%% file: CMET-mdpi.tex
\pdfoutput=1

\PassOptionsToPackage{inline}{enumitem}

\documentclass[preprints,article,accept,moreauthors,pdftex,10pt,a4paper]{mdpi}
\firstpage{1} 
\pubvolume{xx}
\issuenum{1}
\articlenumber{1}
\pubyear{2018}
\copyrightyear{2018}
\externaleditor{Academic Editor: name}
\history{Received: date; Accepted: date; Published: date}

\RequirePackage{etoolbox}
\providetoggle{extendedVersion}
\toggletrue{extendedVersion}
\togglefalse{extendedVersion}
\providetoggle{ieeeVersion}
\toggletrue{ieeeVersion}
\providetoggle{graphicalSummary}
\toggletrue{graphicalSummary}
\togglefalse{graphicalSummary}

\graphicspath{{../figs/}{../pics/}}
\usepackage[caption=false,font=footnotesize]{subfig}

\usepackage{url}

\usepackage{pgfplots} 
\pgfplotsset{compat=1.14} 
\usepgflibrary{shapes.misc} 
\RequirePackage{tikz}
\usetikzlibrary{arrows, shapes,external, decorations.pathmorphing,
    decorations.pathreplacing, backgrounds, positioning, fit, pgfplots.ternary, pgfplots.units}

\Title{Assessing Information Transmission in Data Transformations with the \\ Channel Multivariate Entropy Triangle}

\Author{Francisco J.~Valverde-Albacete$^{1,\ddagger}$\orcidA{}, Carmen~Pel\'aez-Moreno$^{2,\ddagger}$*\orcidB{}}

\AuthorNames{Francisco J.~Valverde-Albacete and Carmen~Pel\'aez-Moreno}

\address{%
$^{1}$ \quad Department of Signal Theory and Communications, Universidad Carlos III de Madrid, Legan\'es 28911, Spain; fva@tsc.uc3m.es\\
$^{2}$ \quad Department of Signal Theory and Communications, Universidad Carlos III de Madrid, Legan\'es 28911, Spain; carmen@tsc.uc3m.es}

\corres{Correspondence: carmen@tsc.uc3m.es; Tel.: +34-91-624-8771}

\firstnote{These authors contributed equally to this work.}

\abstract{%
Data transformation, e.g. feature transformation and selection, is an integral part of any machine learning procedure. %
In this paper we 
introduce an information-theoretic model 
and tools to assess the quality of data transformations in machine learning tasks. 
In an unsupervised fashion, 
we analyze the transfer of information of the transformation of a discrete, multivariate source of information $\overline X$ into a discrete, multivariate sink of information $\overline Y$ related by a distribution $P_{\overline X \overline Y}$\,.
The first contribution is a decomposition of the maximal potential entropy of $(\overline X, \overline Y)$ that we call a \emph{balance equation}, into its a) non-transferable, b) transferable but not transferred and c) transferred parts. 
Such balance equations can  be represented in (de Finetti) entropy diagrams, our second set of  contributions. 
The most important of these, the aggregate Channel Multivariate Entropy Triangle is a visual exploratory tool to assess the effectiveness of multivariate data transformations in transferring information from input to output variables. 
We also show how these decomposition and balance equation also apply to
the entropies of $\overline X$ and $\overline Y$ respectively and generate entropy triangles for them.
As an example, we present the application of these tools %
to the assessment of information transfer efficiency for PCA and ICA as unsupervised feature transformation and selection procedures in supervised classification tasks. %
}


\begin{document}

\section{Introduction}
\label{sec:intro}
Information-related considerations are often cursorily invoked in many machine learning applications sometimes to suggest why a system or procedure is seemingly better than another at a particular task.
In this paper we set out to ground on measurable evidence phrases such as ``this transformation retains more information from the data'' or ``this learning method uses better the information from the data than this other.'' 
%
\input{introCMET.tex}

\section{Methods}
\label{sec:methods}
\input{methodsCMET_R1.tex}

\section{Results}
\label{sec:results}
\input{resultsCMET_R1.tex}

\section{Conclusions}
\label{sec:conc}
\input{conclusionsCMET.tex}

\vspace{6pt} 




\authorcontributions{Conceptualization, Francisco J Valverde-Albacete and Carmen Pel\'aez-Moreno; Formal analysis, Francisco J Valverde-Albacete and Carmen Pel\'aez-Moreno; Funding acquisition, Carmen Pel\'aez-Moreno; Investigation, Francisco J Valverde-Albacete and Carmen Pel\'aez-Moreno; Methodology, Francisco J Valverde-Albacete and Carmen Pel\'aez-Moreno; Software, Francisco J Valverde-Albacete; Supervision, Carmen Pel\'aez-Moreno; Validation, Francisco J Valverde-Albacete and Carmen Pel\'aez-Moreno; Visualization, Francisco J Valverde-Albacete and Carmen Pel\'aez-Moreno; Writing – original draft, Francisco J Valverde-Albacete and Carmen Pel\'aez-Moreno; Writing – review \& editing, Francisco J Valverde-Albacete and Carmen Pel\'aez-Moreno.}

\funding{This research was funded by he Spanish Government-MinECo projects
TEC2014-53390-P 
and 
TEC2017-84395-P 
}


\conflictofinterest{The authors declare no conflict of interest.}

\abbreviations{The following abbreviations are used in this manuscript:\\

\noindent 
\begin{tabular}{@{}ll}
PCA & Principal Component Analysis\\
ICA & Independent Component Analysis\\
CMET & Channel Multivariate Entropy Triangle\\
CBET & Channel Binary Entropy Triangle\\
SMET & Source Multivariate Entropy Triangle
\end{tabular}}

\externalbibliography{yes}
\bibliography{CMET}


\pagebreak
\iftoggle{graphicalSummary}{ %
\begin{figure}
\subfloat[Tranformation block $\overline Y = f(\overline X)$ with $(\overline X, \overline Y)\sim P_{\overline X \overline Y}$]{
	\resizebox{0.6\columnwidth}{!}{
	\input{transformationChain.tex}
	}
}
\subfloat[Entropy diagram for $P_{\overline X \overline Y}$%
]{
	\resizebox{0.4\columnwidth}{!}{
		\input{mod_idiagram_color_multivariate.tex}
		}
}
\\
 \subfloat[Schematic CMET for $P_{\overline X \overline Y}$ with formal interpretation.]{
 	\resizebox{0.6\linewidth}{!}{
  		\input{annotated_triangle_CMET_agg_formal.tex}
  		}
  	}
  		\subfloat[CMET of $\overline Y = f(\overline X)$  for ICA, PCA and log on Iris]{
		\includegraphics[width=0.5\linewidth]{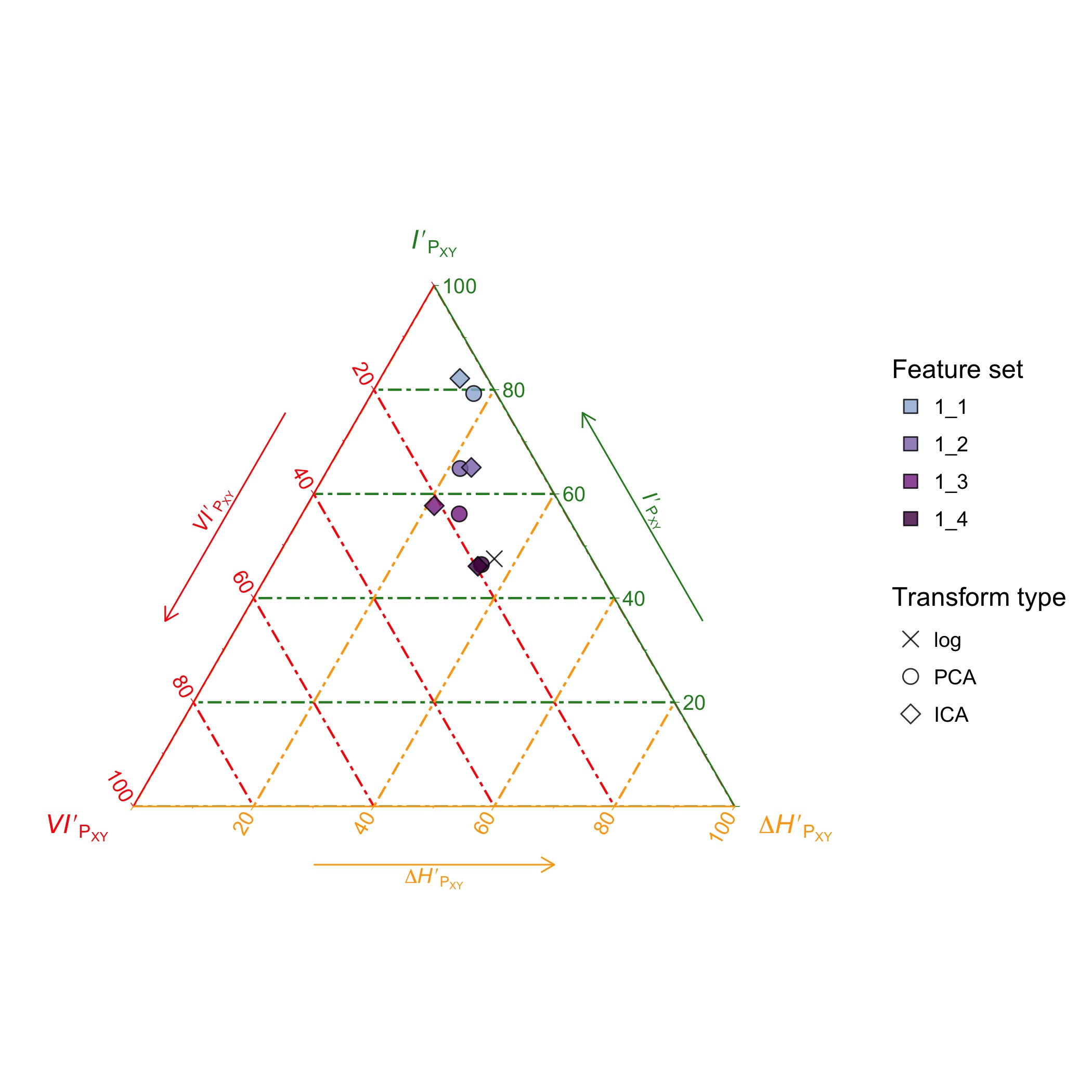}
}

\caption{GRAPHICAL ABSTRACT
}
\end{figure}

}{}%
\end{document}

%% file: introCMET.tex

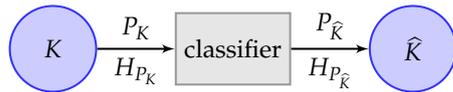
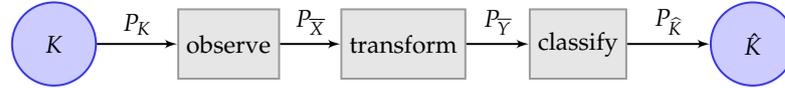
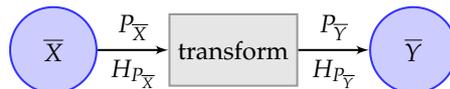
\begin{figure}
\centering
\subfloat[
The end-to-end view for evaluation:  a ``classifier chain'' is trained to predict labels $\widehat K$ from the true emitted labels $K$.]{
      	\resizebox{0.7\linewidth}{!}{
		\input{classifierChain.tex}
		}
\label{fig:multiclass}
}

\vspace{0.5cm}
\subfloat[Conceptual representation of a supervised classification architecture as a communication channel (modified from \protect\cite{val:pel:17b}).]{
\centering
      	\resizebox{0.7\linewidth}{!}{
		\input{multiclassification_channel.tex}
		}
\label{fig:mc:schemes:basic}
}
\\ 
\vspace{0.5cm}
\subfloat[Focusing on the tranformation block implementing $\overline Y = f(\overline X)$. $\overline X$ becomes the data source and $\overline Y$ the sink.]{
\centering
	\resizebox{0.7\columnwidth}{!}{
	\input{transformationChain.tex}
	\vspace{-1.5cm}
	}
\label{fig:transfo}
}
\caption{\textbf{Different views on a supervised classification task as an information channel:} %
\protect\subref{fig:multiclass} for end-to-end evaluation;
\protect\subref{fig:mc:schemes:basic} as individualized blocks;  and 
\protect\subref{fig:transfo} focused on the transformation. 
\label{fig:it:schemes}
}
\end{figure}

{
This has become particularly relevant with the increase of complexity of machine learning methods, such as deep neuronal architectures~\cite{goo:ben:cou:16}, that prevents straightforward interpretations. Nowadays, these learning schemes are almost always becoming \textit{black-boxes} where the researchers  try to optimize a prescribed performance metric without looking inside. 
However, there is a need to assess what are the deep layers actually accomplishing. 
Although some answers start to appear~\cite{sch:tis:17,tis:zas:15}, the issue is by no means settled. 

In this paper, we put forward that framing the previous problem into a generic information-theoretical model can shed light onto it by exploiting the versatility of Information Theory. 
For instance, a classical end-to-end example of an information-based model evaluation can be observed in Figure ~\ref{fig:it:schemes}.\subref{fig:multiclass}. In this supervised scheme introduced in~\cite{val:pel:14a}, the evaluation of the performance of the classifier involves only the comparison of the true labels $K$ vs. the predicted labels $\hat K$. This means that all the complexity enclosed in the \textit{classifier} box cannot be accessed, measured or interpreted.

In this paper, we want to expand the previous model into the scheme of 
Figure~\ref{fig:it:schemes}.\subref{fig:mc:schemes:basic} that provides a more detailed picture of the contents of the \textit{black-box} where: %
}

{

\begin{itemize}
\item  A random source of \emph{classification labels} $K$ is subjected to a measurement process that returns random \emph{observations} $\overline X$. 
The $n$ instances of pairs $(k_i, \overline x_i), 1 \leq i \leq n$ is often called the \emph{(task) dataset}. 

\item Then a generic \emph{data transformation} block 
may transform the available data---e.g. the observations in the dataset $\overline X$---into another data with ``better'' characteristics---the transformed \emph{feature vectors} $\overline Y$\,. 
These characteristics may be representational power, independence between individual dimensions, reduction of complexity offered to a classifier, etc. The process is normally called \emph{feature transformation and selection}. 
%
%

\item Finally, the $\overline Y$  are the inputs to an actual  classifier of choice that obtains the \emph{predicted  labels} $\hat K$. 
\end{itemize}
This would allow us to better understand the flow of information in the classification process with a view to assessing and improving it. 

Note the similarity between the classical setting of Figure ~\ref{fig:it:schemes}.\subref{fig:multiclass} and the transformation block of Figure  ~\ref{fig:it:schemes}.\subref{fig:mc:schemes:basic} reproduced in Figure ~\ref{fig:it:schemes}.\subref{fig:transfo} for convenience. Despite this,  the former represents a single-input single-output  (SISO) block with $(K,{\hat K}) \sim P_{K{\hat K}}$  
whereas the later represents a multivariate multiple-input multiple-output (MIMO) block described by the joint distribution of random vectors $(\overline X, \overline Y) \sim P_{\overline X\overline Y}$.  

This MIMO kind of block may represent an  unsupervised transformation method---for instance, a Principal Component Analysis (PCA) or Independent Component Analysis (ICA)---in which case the ``effectiveness'' of the transformation is supplied by a heuristic principle, e.g. least reconstruction error on some test data, maximum mutual information, etc. 
But it may also represent a supervised transformation method%
---for instance, $\overline X$ are the feature instances and $\overline Y$ are the (multi)labels or classes in a classification task, 
or $\overline Y$ may be the activation signals of a convolutional neural network trained using an implicit target signal---
in which case, the ``effectiveness'' should measure the conformance to the supervisory signal. 

In ~\cite{val:pel:14a} we argued 
for carrying out the evaluation of classification tasks that can be modeled by Figure~\ref{fig:it:schemes}.\subref{fig:multiclass}  with 
%
the new framework of \emph{entropy balance equations} and their related \emph{entropy triangles}~\citep{val:pel:10b,val:pel:14a,val:pel:17b}. This has provided a means of quantifying and visualizing the end-to-end \textit{information transfer} for SISO architectures. 
%
%
The gist of this framework is explained in Section~\ref{sec:et}: 
if a classifier working on a certain dataset obtained a confusion matrix $P_{K\hat K}$, then 
we can information-theoretically  assess the classifier by analyzing the entropies and informations in the related distribution $P_{K\hat K}$ with the help of a \emph{balance equation}~\citep{val:pel:10b}. 
However, looking inside the \textit{black-box} poses a challenge since $\overline X$ and $\overline Y$\, are random vectors and most information-theoretic quantities are not readily available in their multivariate version.

If we want to extend the same framework of evaluation to random vectors
in general, we need the multivariate generalizations of the information-theoretic measures 
involved in the \textit{balance equations}, an issue that is not free of contention.  
With this purpose in mind, we  review the best-known multivariate generalizations of mutual information in Section~\ref{sec:circa:MI}. 
We present our contributions finally in Section~\ref{sec:results}. 
As a first result we develop a balance equation for the joint distribution $P_{\overline X \overline Y}$ and related representation in Sections~\ref{sec:theory} and~\ref{sec:visual}, respectively. But we are also able to obtain split equations for the input and output multivariate sources only tied by one multivariate extension of mutual information, much as in the SISO case. 
As an instance of use,  in Section~\ref{sec:app} we analyze the transfer of information in PCA and ICA transformations applied to some well-known UCI datasets. 
We conclude with a discussion of the tools in light of this application in Section~\ref{sec:discuss}. 

}

%% file: classifierChain.tex
\tikzstyle{source_node}=[circle,
                   thick,
                   minimum size=1.2cm,
                   draw=blue!80,
                   fill=blue!20]
\tikzstyle{empty_node}=[rectangle]
\tikzstyle{block_node}=[rectangle,
                   thick,
                   minimum size=1cm,
                   draw=gray!80,
                   fill=gray!20]
\begin{tikzpicture}[node distance=2.5cm, auto,>=latex', thick]
    \path[use as bounding box] (-1,0) rectangle (10,-1);
    \path[->] node[source_node] (source) {$K$};
    \path[->] node[block_node, right of=source] (transfo) {classifier}
    			    (source) edge node {$P_{K}$} node[midway,below] {$H_{P_{K}}$} (transfo);
    \path[->] node[source_node, right of=transfo] (sink) {$\widehat K$}			    
    			    (transfo) edge node {$P_{\widehat K}$} node[midway,below] {$H_{P_{\widehat K}}$} (sink);
\end{tikzpicture}

%% file: multiclassification_channel.tex
%
%
\tikzstyle{format} = [draw, thin, fill=blue!20]
\tikzstyle{medium} = [ellipse, draw, thin, fill=green!20, minimum height=2.5em]
\tikzstyle{bubble} = [circle, draw, thin]
\tikzstyle{block} = [draw, thin]
\tikzstyle{source_node}=[circle,
                   thick,
                   minimum size=1.2cm,
                   draw=blue!80,
                   fill=blue!20]
\tikzstyle{empty_node}=[rectangle]
\tikzstyle{block_node}=[rectangle,
                   thick,
                   minimum size=1cm,
                   draw=gray!80,
                   fill=gray!20]

\begin{tikzpicture}[node distance=2.5cm, auto,>=latex', thick]
    \path[use as bounding box] (-1,0) rectangle (10,-1);
    \path[->] node[source_node] (source) {$K$};
    \path[->] node[block_node, right of=source] (encoder) {observe}
			    (source) edge node {$P_K$} (encoder);
    \path[->] node[block_node, right of=encoder] (channel) {transform}
			    (encoder) edge node {$P_{\overline X}$} (channel);
    \path[->] node[block_node, right of=channel] (decoder) {classify}
			    (channel) edge node {$P_{\overline Y}$} (decoder);
    \path[->] node[source_node, right of=decoder] (sink) {$\hat K$}			    
			    (decoder) edge node {$P_{\widehat K}$} (sink);
\end{tikzpicture}

%% file: transformationChain.tex
\tikzstyle{source_node}=[circle,
                   thick,
                   minimum size=1.2cm,
                   draw=blue!80,
                   fill=blue!20]
\tikzstyle{empty_node}=[rectangle]
\tikzstyle{block_node}=[rectangle,
                   thick,
                   minimum size=1cm,
                   draw=gray!80,
                   fill=gray!20]
\begin{tikzpicture}[node distance=2.5cm, auto,>=latex', thick]
    \path[use as bounding box] (-1,0) rectangle (10,-1);
    \path[->] node[source_node] (source) {$\overline X$};
    \path[->] node[block_node, right of=source] (transfo) {transform}
    			    (source) edge node {$P_{\overline X}$} node[midway,below] {$H_{P_{\overline X}}$} (transfo);
    \path[->] node[source_node, right of=transfo] (sink) {$\overline Y$}			    
    			    (transfo) edge node {$P_{\overline Y}$} node[midway,below] {$H_{P_{\overline Y}}$} (sink);
\end{tikzpicture}

%% file: methodsCMET_R1.tex
{ 
In Section~\ref{sec:results} we will build a solution to our problem by finding the minimum common multiple, so to speak, of our previous solutions to the SISO block we describe in Section  \ref{sec:et} and the multivariate source cases, to be described in Section \ref{sec:circa:MI}. 
}
\subsection{The Channel Bivariate Entropy Balance Equation and Triangle}
\label{sec:et}
\input{cbbe_cbetR1.tex}

\subsection{Quantities around the Multivariate Mutual Information}
\label{sec:circa:MI}
\input{circaMutualInformationR1.tex}


%% file: cbbe_cbetR1.tex
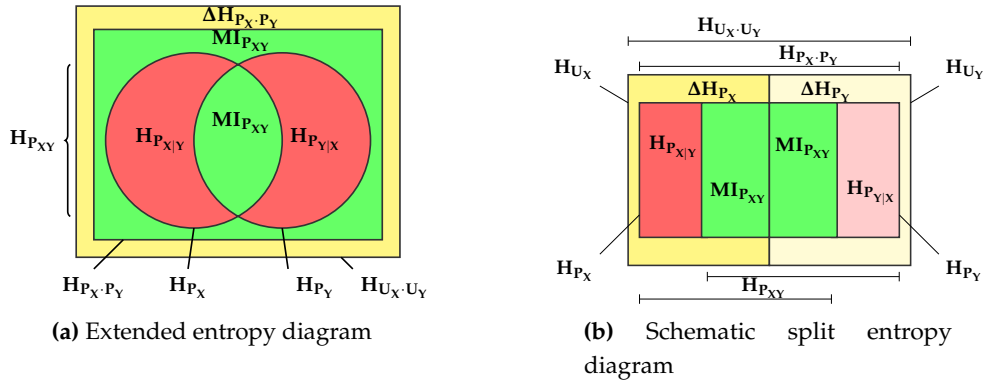
\begin{figure*}[!ht]%
\centering
\subfloat[Extended entropy diagram]{%
	\resizebox{0.4\columnwidth}{!}{
  		\input{mod_idiagram_color.tex}
  		}
  \label{fig:cbet:idiagram}
}%
\hfil%
\subfloat[Schematic split entropy diagram]{%
	\resizebox{0.4\columnwidth}{!}{
  		\input{ext_mod_idiagram_color.tex}
  		}
\label{fig:cbet:schematics}
}%
\caption[]{{\bf Extended entropy diagram related to
    a bivariate distribution, from \citep{val:pel:10b}.}
}%
\label{fig:cbet:diagrams}%
\end{figure*}

A solution to conceptualizing and visualizing the transmission of information through a channel where input and output are reduced to a single variable, that is with $|\overline X| = 1$ and $|\overline Y| = 1$ , was presented in~\cite{val:pel:10b} and later extended in~\cite{val:pel:14a}. 
For this case we use simply $X$ and $Y$ to describe the random variables\footnote{In the introduction, and later in the example application, these were called $K$ and $\hat K$ but here we want to present this case as a simpler version of the one we set out to solve in this paper.} and Figure~\ref{fig:cbet:diagrams}.{\subref{fig:cbet:idiagram}} 
depicts a classical information-diagram (i-diagram)\cite{yeu:91,rez:61} of an entropy decomposition around $P_{XY}$ to which we have included the exterior boundaries arising from \textit{entropy balance equation} as we will show later. Three
crucial regions can be observed:
{ 
\begin{itemize}
\item The \emph{(normalized) redundancy~\cite[$\S~2.4$]{kay:03}, or divergence with respect to uniformity} (yellow area), $\Delta H_{P_X \cdot P_Y}$, between the joint
distribution where $P_X$ and $P_Y$ are independent and the uniform distributions with the
same cardinality of events as $P_X$ and $P_Y$\,,
\begin{align}
\label{eq:cbet:delta}
\Delta H_{P_X\cdot P_Y} = H_{U_X\cdot   U_Y}- H_{P_X \cdot P_Y}\,.
\end{align}

\item The \emph{mutual information}, $MI_{P_{XY}}$~\cite{sha:48a,sha:48b} (each of the green areas), quantifies the force of the stochastic binding between $P_X$ and $P_Y$\,, ``towards the outside'' in Fig.~\ref{fig:cbet:diagrams},\subref{fig:cbet:idiagram}
\begin{align}
    \label{eq:cbet:mi}
MI_{P_{XY}} = H_{P_X \cdot P_Y} - H_{P_{XY}}
\end{align}
but also ``towards the inside'', 
\begin{align}
    \label{eq:cbet:mi:alt}
	MI_{P_{XY}}  = H_{P_{\overline X}} - H_{P_{\overline X| \overline Y}}= H_{P_{\overline Y}} - H_{P_{\overline Y| \overline X}}\,.
\end{align}
\item The \emph{variation of information} (the sum of the red areas), $VI_{P_{XY}}$~\cite{mei:07}, embodies 
the residual entropy, not used in binding the  variables, 
\begin{align}
    \label{eq:cbet:vi}
    VI_{P_{XY}} = H_{P_{X|Y}}+H_{P_{Y|X}}\,.
\end{align}
\end{itemize}
}

Then, we may write the following \emph{entropy balance equation} between the entropies of $X$ and $Y$: 
\begin{align}
  \label{eq:cbet:balance}
H_{U_X\cdot U_Y} &= \Delta H_{P_X\cdot P_Y} + 2* MI_{P_{XY}} +  VI_{P_{XY}} 
  \\
  0 &\leq \Delta H_{P_X\cdot P_Y}, MI_{P_{XY}},  VI_{P_{XY}} \leq  H_{U_X\cdot U_Y}\notag
\end{align}
where the bounds are easily obtained from distributional considerations~\cite{val:pel:10b}.
If we normalize \eqref{eq:cbet:balance} by the overall entropy $H_{U_X \cdot U_Y}$ we
obtain 
\begin{align}
\label{eq:cbet:simplex}
  1 &= \Delta' H_{P_X\cdot P_Y} + 2* MI'_{P_{XY}} +  VI'_{P_{XY}}
  &
  0 &\leq \Delta' H_{P_X\cdot P_Y}, MI'_{P_{XY}},  VI'_{P_{XY}} \leq  1 
\end{align}

Equation~\eqref{eq:cbet:simplex} is the $2$-simplex in normalized $\Delta {H'}_{P_X \cdot P_Y} \times 2 {\emph{MI}\,'}_{P_{XY}} \times {\emph{VI}\,'}_{P_{XY}}$ space. 
Each joint distribution $P_{XY}$ can be characterized by its \emph{joint entropy fractions}, $F(P_{XY}) = [\Delta H'_{P_{XY}},2\times \emph{MI}\:'_{P_{XY}},\emph{VI}\:'_{P_{XY}}]$\,,
%
whose projection onto the plane with director vector $(1,1,1)$ is its \emph{de Finetti  or Compositional diagram~\cite{paw:ego:tol:15}}. 
This diagram of the 2-simplex is an equilateral triangle whose coordinates are  $F(P_{XY})$ 
so {every bivariate distribution shows as a point in the triangle}, and each zone in the triangle is indicative of the characteristics of distributions whose coordinates fall in it.  This is what we call the Channel Bivariate Entropy Triangle, CBET, an schematic of which 
is shown in Fig.~\ref{fig:cbet:classification}. 

Considering  \eqref{eq:cbet:balance} and the composition of the quantities in it we can actually decompose the equation into two \emph{split balance} equations, 
\begin{align}
  \label{eq:cbet:split:balance}
  H_{U_{X}} &= \Delta H_{P_X} + \emph{MI}_{P_{XY}} + H_{P_{X|Y}} & \quad
  H_{U_{Y}} &= \Delta H_{P_Y} + \emph{MI}_{P_{XY}} + H_{P_{Y|X}}\,.
\end{align}
with the obvious limits.
%
These can be each normalized by $H_{U_{X}}$, respectively $H_{U_{Y}}$,
leading to the 2-simplex equations 
\begin{align}
  \label{eq:cbet:split:simplex}
  1 &= \Delta'H_{P_X} + \emph{MI'}_{P_{XY}} + H'_{P_{X|Y}} & \quad
  1 &= \Delta'H_{P_Y} + \emph{MI}'_{P_{XY}} + H'_{P_{Y|X}}\,.
\end{align}
%
Since these are also equations on a 2-simplex, we can actually represent the coordinates $F_X(P_{XY}) = [\Delta H'_{P_{X}}, \emph{MI}\:'_{P_{XY}},H'_{P_{X|Y}} ]$ and $F_Y(P_{XY}) = [\Delta H'_{P_{Y}},\emph{MI}\:'_{P_{XY}},H'_{P_{Y|X}}]$ in the same triangle side by side the original $F(P_{XY})$, whereby the representation seems to split in two. 

\subsubsection{Application: the evaluation of multiclass classification}
The CBET can be used to visualize the performance of supervised classifiers in a straightforward manner as announced in the introduction:  
consider the confusion matrix $N_{K\hat{K}}$ of a classifier chain on a supervised classification task given the random variable of true class labels $K \sim P_K$ and that of predicted labels $\widehat K \sim P_{\widehat K}$ as depicted in Figure~\ref{fig:it:schemes}.\subref{fig:multiclass}---that now play the role of $P_X$ and $P_Y$. 
From this confusion matrix we can estimate the 
 joint distribution $P_{K\widehat K}$ between the random variables, so that  the entropy triangle for  $P_{K\widehat K}$ produces valuable information about the actual classifier used to solve the task~\cite{val:pel:10b,val:car:pel:13}, and even the theoretical limits of the task---for instance, whether it can be solved in a trustworthy manner by classification technology, and with what effectiveness. 

\begin{figure}
\centering
\input{annotated_triangle_CBET_multiclass.tex}
\caption{\textbf{Schematic CBET as applied to supervised classifier assessment.} An actual triangle shows dots for each classifier (or its split coordinates, see Fig. \ref{fig:transfo:independent} for example) and none of the callouts for specific types of classifiers (from~\protect\citep{val:pel:14a}). The callouts situated in the center of the sides of the triangle apply to the whole side.
\label{fig:cbet:classification}
}
\end{figure}
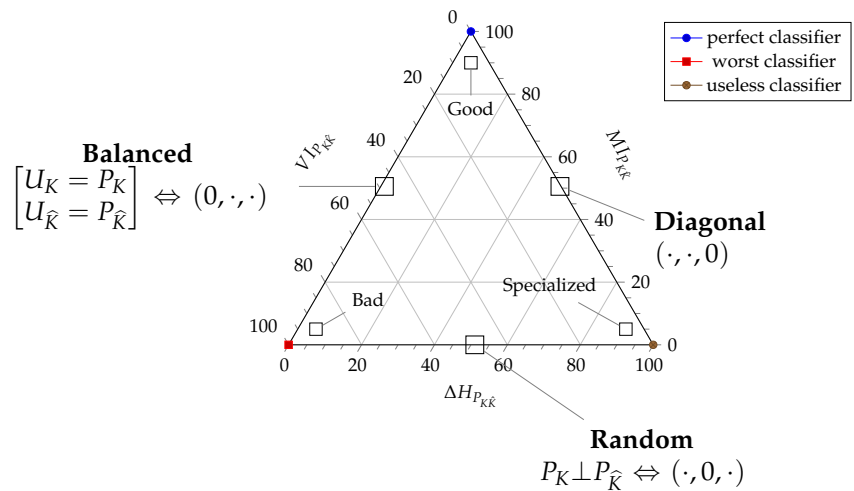

The CBET acts, in this case, as an exploratory data analysis tool for visual assessment, as shown in Figure~\ref{fig:cbet:classification}. 
The success of this approach in the bivariate, supervised classification case is a strong hint that the multivariate extension will likewise be useful for other machine learning tasks. See~\cite{val:pel:14a} for a thorough explanation of this procedure. 


%% file: mod_idiagram_color.tex
\def\firstcircle{(0,0) circle (1.5cm)}
\def\secondcircle{(0:1.5cm) circle (1.5cm)}

\colorlet{circle edge}{black!80}
\colorlet{circle area}{red!60}

\tikzset{filled/.style={fill=circle area, draw=circle edge, thick},
    outline/.style={draw=circle edge, thick}}
\tikzset{bold/.style={font={\bfseries}}}

\begin{tikzpicture}[thick,even odd rule]
    \draw[filled,color=yellow!60,outline=black!80] (-2cm,-2cm) rectangle (3.5cm,2.3cm);
    \draw[filled,color=green!60,outline=black!80] (-1.7cm,-1.7cm) rectangle (3.2cm,1.9cm);
    \draw[filled]
           \firstcircle node[anchor=east] {$\mathbf{H_{P_{X|Y}}}$}
           \secondcircle node[anchor=west] {$\mathbf{H_{P_{Y|X}}}$};
    \node[anchor=south,bold] at (0.8cm,0) {$\mathbf{MI_{P_{XY}}}$};
    \node[anchor=south] at (0.8cm,1.4cm) {$\mathbf{MI_{P_{XY}}}$};
    \node
         at (0.8cm,2.1cm) {$\mathbf{\Delta H_{P_X\cdot P_Y}}$};
    \draw (2.5,-2)   --  (2.7cm,-2.2cm) node[anchor=north west]
    {$\mathbf{H_{U_X\cdot U_Y}}$}; 
   \node[anchor=east] at (-2.2cm,0cm) {$\mathbf{H_{P_{XY}}}$}; 
    \draw[decorate,decoration=brace] (-2.1,-1.3) -- (-2.1,1.3){};
    \draw (-1.2,-1.7) --  (-1.7cm,-2.2cm)  node[anchor=north]
        {$\mathbf{H_{P_{X} \cdot P_{Y}}}$}; 
    \draw (0,-1.5) --  (-0.1cm,-2.2cm)  node[anchor=north] {$\mathbf{H_{P_X}}$}; 
    \draw (1.5,-1.5) --  (1.6cm,-2.2cm)  node[anchor=north west ] {$\mathbf{H_{P_Y}}$}; 
  \end{tikzpicture}

%% file: ext_mod_idiagram_color.tex
\colorlet{circle edge}{black!80}
\colorlet{circle area}{red!60}

\tikzset{filled/.style={fill=circle area, draw=circle edge, thick},
    outline/.style={draw=circle edge, thick}}
\tikzset{bold/.style={font={\bfseries}}}

  \begin{tikzpicture}[very thick,even odd rule]
    \draw (0,1.7) -- (0,-1.7);
    \draw [filled,color=yellow!60,outline=black!80] (-2.5,-1.7) rectangle (0,1.7);
    \draw [filled,color=yellow!20,outline=black!80] (0,-1.7) rectangle (2.5,1.7);
    \draw [filled,color=red!60,outline=black!80](-2.3,-1.2) rectangle (-1.1,1.2);
    \draw [filled,color=red!20,outline=black!80](1.1,-1.2) rectangle (2.3,1.2);
    \draw [filled,color=green!60,outline=black!80](-1.2,1.2) rectangle (0,-1.2);
    \draw [filled,color=green!60,outline=black!80](0,1.2) rectangle (1.2,-1.2);
    \begin{scope}[ultra thin]
      \draw(-1.7,0.4) node
                           {$\mathbf{H_{P_{X|Y}}}$};
      \draw(-0.55,-0.4) node
                           {$\mathbf{MI_{P_{XY}}}$};
      \draw(0.6,0.4) node
                           {$\mathbf{MI_{P_{XY}}}$};
      \draw(1.8,-0.4) node
                           {$\mathbf{H_{P_{Y|X}}}$};
      \draw(-1.0,1.4) node
                           {$\mathbf{\Delta H_{P_X}}$};
      \draw(1.0,1.4) node
                           {$\mathbf{\Delta H_{P_Y}}$};
      \draw (2.5, 1.2) -- (3.0, 1.5) node
                           [anchor=south west] 
                           {$\mathbf{H_{U_Y}}$};
      \draw (2.3, -0.8) -- (3.0, -1.5) node
                           [anchor=north west] 
                           {$\mathbf{H_{P_Y}}$};
      \draw (-2.5, 1.2) -- (-3.0, 1.5) node
                           [anchor=south east] 
                           {$\mathbf{H_{U_X}}$};
      \draw (-2.3, -0.8) -- (-3.0, -1.5) node
                           [anchor=north east] 
                           {$\mathbf{H_{P_X}}$};
      \draw[|-|](-2.5,2.3) -- (2.5,2.3);
      \draw(-0.7,2.55) node {$\mathbf{H_{U_X \cdot U_Y}}$};
      \draw[|-|](-2.3,1.85) -- (2.3,1.85);
      \draw(0.7,2.05) node {$\mathbf{H_{P_X \cdot P_Y}}$};
      \draw[|-|](-1.1,-1.9) -- (2.3,-1.9);
      \draw[|-|](-2.3,-2.3) -- (1.1,-2.3);
      \draw(-0.1,-2.1) node {$\mathbf{H_{P_{XY}}}$};
    \end{scope}
  \end{tikzpicture}

%% file: annotated_triangle_CBET_multiclass.tex

\begin{tikzpicture}[scale=0.7]

\begin{ternaryaxis}[
 xmin=0,
 xmax=100,
 ymin=0,
 ymax=100,
 zmin=0,
 zmax=100, 
 xlabel=$MI_{P_{K\hat K}}$,
 ylabel=$VI_{P_{K\hat K}}$,
 zlabel=$\Delta H_{P_{K\hat K}}$,
 label style={sloped},
 minor tick num=3,
 grid=major
]
\addplot3 table {
  M V D
  100 0 0
};
\addlegendentry{perfect classifier}
\addplot3 table {
  M V D
  0 100 0
};
\addlegendentry{worst classifier}
\addplot3 table {
  M V D
  0 0 100
};
\addlegendentry{useless classifier}

\node[pin=270:Good,draw=black] at (axis cs:90,5) {};
\node[pin=130:Specialized,draw=black] at (axis cs:5,5) {};
\node[pin=30:Bad,draw=black] at (axis cs:5,90) {};

\end{ternaryaxis}

\node[pin={[pin distance=1cm,text width=3cm,align=center] 
  300:\textbf{Random $P_K \bot P_{\widehat K} \Leftrightarrow (\cdot,0,\cdot)$ }}, draw=black]  
   at (3.5,0) {};P
\node[pin={[pin distance=1cm,text width=1cm,align=center] 350:\textbf{Diagonal $  (\cdot,\cdot,0)$ }}, draw=black] at (5.1,3) {};
\node[pin={[pin distance=1cm,text width=4cm,align=center] 180:%
\textbf{Balanced }
$
    \begin{bmatrix}
      U_K = P_K \\ U_{\widehat{K}} = P_{\widehat{K}}      
    \end{bmatrix}
  \Leftrightarrow (0,\cdot,\cdot)$}, draw=black] at (1.8,3) {};

\end{tikzpicture}







%% file: circaMutualInformationR1.tex
\label{sec:circaMI}
The main hurdle for a multivariate extension of the balance equation~\eqref{eq:cbet:balance} and the CBET is the multivariate generalization of binary mutual information,  
since it quantifies the information transport from input to output in  the bivariate case, and is also crucial for the decoupling of~\eqref{eq:cbet:balance} into the split balance equations~\eqref{eq:cbet:split:balance}.
%
For this reason, we next review the different ``flavors'' of information measures describing sets of more than two variables looking for these two properties. We start from very basic definitions both in the interest of self-containment and to provide a script on the process of developing future analogues  for other information measures. 

To fix notation, let $\overline X = \{X_i \mid 1 \leq i \leq n\}$ be a set of discrete random variables with joint multivariate distribution $P_{\overline X} = P_{X_1\ldots{X_n}}$, and the corresponding marginals $P_{X_i} (x_i) = \sum_{j\neq i}P_{\overline X}(\overline x)$ where $\overline x = x_1\ldots{x_n}$ is a tuple of $n$ elements. 
And likewise for $\overline Y = \{Y_j \mid 1 \leq j \leq l\}$, with $P_{\overline Y}=P_{Y_1\ldots{Y_l}}$ and the marginals $P_{Y_j}$\,. 
Furthermore let $P_{\overline X \overline Y}$ be the joint distribution of the $(n + l)$-length tuples $\overline X\overline Y$\,. 

{ 
\noindent Note that two different \textit{Situations} can be clearly distinguished:
\begin{enumerate}[leftmargin=2cm,labelsep=2mm]
\item [Situation 1:] all the random variables form part of the same set $\overline X$ and we are looking at information transfer \emph{within} this set, or 

\item [Situation 2:] are partitioned into two different sets $\overline X$ and $\overline Y$ and we are looking at information transfer \emph{between} these sets. 

\end{enumerate}
}
An up-to-date review of multivariate information measures in both situations is~\cite{tim:alf:fle:beg:14} that follows 
the interesting methodological point from~\cite{jam:ell:cru:11}  of calling \emph{information} those measures which involve amounts of entropy shared by multiple variables and \emph{entropies} those that do not\footnote{Although this poses a conundrum for the entropy  written as the self information $H_{P_{X}} = MI_{P_{XX}}$.%
}. 

Since i-diagrams 
are a powerful tool to visualize the interaction of distributions in the bivariate case, we will also try to use them for sets of random variables. 
%
%
For multivariate generalizations of mutual information as seen in the i-diagrams, the following caveats apply:
\begin{itemize}
\item  Their multivariate generalization is only warranted when signed measures of probability are considered, since it is well-known that some of these ``areas'' can be \emph{negative}, contrary to geometric intuitions on this respect. 

\item We should retain the bounding rectangles that appear when considering the most entropic distributions with similar support to the ones being graphed~\cite{val:pel:10b}. This is the sense of the bounding rectangles in Figures~\ref{fig:midiagrams}.\subref{fig:smet:idiagram} and~~\ref{fig:midiagrams}.\subref{fig:cmet:idiagram}.
\end{itemize}
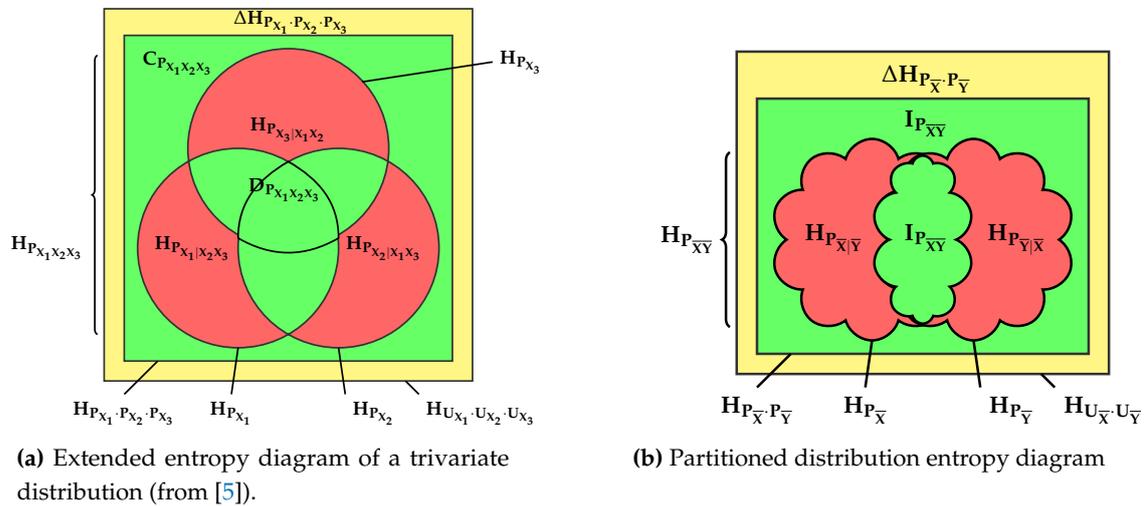
\begin{figure*}[!ht]%
  \centering
\subfloat[{Extended entropy diagram of 
    a trivariate distribution (from~\protect\cite{val:pel:17b}).}
]{
	\resizebox{0.5\columnwidth}{!}{
  		\input{multi_mod_idiagram_color_barX.tex}
  		}
  \label{fig:smet:idiagram}
}%
\subfloat[Partitioned distribution entropy diagram%
]{
	\resizebox{0.5\columnwidth}{!}{
		\input{mod_idiagram_color_multivariate.tex}
		}
  \label{fig:cmet:idiagram}
}
\caption[]{(Color Online) 
{\bf Extended entropy diagram of multivariate distributions for 
	\protect\subref{fig:smet:idiagram} a trivariate distribution (from~\protect\cite{val:pel:17b}) as an instance of \textit{Situation} 1, and
	\protect
	\subref{fig:cmet:idiagram} a joint distribution where a partitioning of the variables is made evident (\textit{Situation} 2).}
The color scheme follows that of Fig.~\protect\ref{fig:cbet:diagrams}, to be explained in the text. 
}%
\label{fig:midiagrams}%
\end{figure*}

With great insight, the authors of~\cite{jam:ell:cru:11} point out that some of the multivariate information measures stem from focusing in a particular property of the bivariate mutual information and generalize it to the multivariate setting.
The properties in question are:
\begin{align}
MI_{P_{XY}} &= H_{P_{X}} + H_{P_{Y}} - H_{P_{XY}} \tag{\ref{eq:cbet:mi}}
\\
MI_{P_{XY}} &= H_{P_X} - H_{P_{X|Y}} = H_{P_Y} - H_{P_{Y|X}} \tag{\ref{eq:cbet:mi:alt}}
\\
\label{prop:mi3}
MI_{P_{XY}} &= \sum_{x, y} P_{XY}(x,y) \log\frac{P_{XY}(x, y)}{P_{X}(x)P_{Y}(y)}
\end{align}
Regarding the first situation of a vector of random variables $\overline X \sim P_{\overline X}$, let $\Pi_{\overline X} = \prod_{i=1}^{n} P_{X_i}$ be the (jointly) independent distribution with similar marginals to $P_{\overline X}$. To picture this (virtual) distribution
consider Figure~\ref{fig:midiagrams}.\subref{fig:smet:idiagram} depicting an i-diagram for $\overline X = [X_1, X_2, X_3]$. Then $\Pi_{\overline X} = P_{X_1}\cdot P_{X_2} \cdot P_{X_3}$ is the inner rectangle containing both green areas. 
The different extensions of mutual information that concentrate on different properties are:
\begin{itemize}
\item the \emph{total correlation}~\cite{wat:60}, \emph{integration}~\cite{ton:spo:ede:94} or \emph{multiinformation}~\cite{stu:vej:98} which is a generalization of \eqref{eq:cbet:mi}, represented by the green area outside $H_{P_{\overline X}}$.
\begin{align}
\label{def:tc}
C_{P_{\overline X}} = H_{\Pi_{\overline X}} - H_{P_{\overline X}}
\end{align}
\item the \emph{dual total correlation}~\cite{han:78,abd:plu:12} or \emph{interaction complexity}~\cite{ton:ede:spo:98} is a generalization of \eqref{eq:cbet:mi:alt}, represented by the green area inside $H_{P_{\overline X}}$
\begin{align}
\label{def:dtc}
D_{P_{\overline X}} = H_{P_{\overline X}} - VI_{P_{\overline X}}
\end{align}
\item the \emph{interaction information}~\cite{mcg:54}, \emph{multivariate mutual information}~\cite{han:80}  or \emph{co-information}~\cite{bell:03} is the generalization of \eqref{prop:mi3}, the total amount of information to which all variables contribute. 
\begin{align}
\label{def:mmi}
MI_{P_{\overline X}} = \sum P_{\overline X}(\overline x) \log\frac{P_{\overline X}(\overline x)}{\Pi_{\overline X}(\overline x)}
\end{align}
It is represented by the inner convex green area (within the dual total correlation), but 
note that it may in fact be negative for $n > 2$\,\cite{Abdallah:2010te}.

\item the \emph{local exogenous information}~\cite{jam:ell:cru:11} or the \emph{bound information}~\cite{val:pel:16a} is the addition of the total correlation and the dual total correlation
\begin{align}
\label{eq:multiM}
M_{P_{\overline X}} &= C_{P_{\overline X}} + D_{P_{\overline X}}\,.
\end{align}

\end{itemize}

Some of these generalizations of the multivariate case 
were used in~\cite{val:pel:16a,val:pel:17b} to develop a similar technique as the CBET but applied to analyzing the information content of data sources. 
For this purpose, it was necessary to define for every random variable
a \emph{residual entropy} $H_{P_{X_i\mid X_i^c}}$---where ${X_i^c} = \overline X\setminus\{X_i\}$---%
which is not explained by the information provided by the other variables.
We call \emph{residual information}~\cite{jam:ell:cru:11} or \emph{(multivariate) variation of information}~\cite{mei:07,val:pel:16a} to the generalization of the same quantity in the bivariate case, i.e. the sum of these quantities across the set of random variables:
\begin{align}
\label{eq:smet:vi}
VI_{P_{\overline X}} = \sum_{i=1}^{n} H_{P_{X_i\mid X_i^c}}\,.
\end{align}
Then the variation of information can easily be seen to consist of the sum of the red areas in Figure~\ref{fig:midiagrams}.\subref{fig:smet:idiagram}  and amounts to information peculiar to each variable. 

{ 
The main question regarding this issue is which---if any---of these generalizations of bivariate mutual information are adequate for an analogue of the entropy balance equations and triangles. 
Note that all of these generalizations consider $\overline X$ as a homogeneous set of variables, that is, the \textit{Situation 1} described at the beginning of this section,
and none consider the partitioning of the variables in $\overline X$ into two subsets (\textit{Situation 2}), for instance  
to distinguish between input and output ones, so the answer cannot be straightforward. 
%
This issue is clarified in Section~\ref{sec:theory}.
}

%% file: multi_mod_idiagram_color_barX.tex
\def\firstcircle{(0,0) circle (1.5cm)}
\def\secondcircle{(0:1.5cm) circle (1.5cm)}
\def\thirdcircle{(0.75,1.5) circle (1.5cm)} 
\colorlet{circle edge}{black!80}
\colorlet{circle area}{red!60}

\tikzset{filled/.style={fill=circle area, draw=circle edge, thick},
    outline/.style={draw=circle edge, thick}}
\tikzset{bold/.style={font={\bfseries}}}

\begin{tikzpicture} 
  [thick,scale=1.1,every node/.style={scale=0.9}, even odd rule]
    \draw[filled,color=yellow!60,outline=black!80] (-2cm,-2cm) rectangle (3.5cm,3.6cm);
    \draw[filled,color=green!60,outline=black!80] (-1.7cm,-1.7cm) rectangle (3.2cm,3.2cm);
    \draw[filled]
           \firstcircle node[anchor=east] {$\mathbf{H_{P_{X_1|X_2X_3}}}$}
           \secondcircle node[anchor=west] {$\mathbf{H_{P_{X_2|X_1X_3}}}$}
           \thirdcircle node[anchor=south] {$\mathbf{H_{P_{X_3|X_1 X_2}}}$};
	\fill [fill=green!60,draw=black] (0.0,0.15) to [out=90,in=210] (0.75,1.3) 
		to [out=-30,in=90] (1.5,0.15) 
		to [out=210,in=-30] (0.0,0.15);
    \node[anchor=south,bold] at (0.7cm,0.6cm) {$\mathbf{D_{P_{X_1X_2X_3}}}$};
    \node[anchor=south] at (-0.9cm,2.5cm) {$\mathbf{C_{P_{X_1X_2X_3}}}$};
    \node
         at (0.8cm,3.4cm) {$\Delta\mathbf{H_{P_{X_1}\cdot P_{X_2} \cdot P_{X_3}}}$};
    \draw (2.5,-2)   --  (2.7cm,-2.2cm) node[anchor=north west]
    {$\mathbf{H_{U_{X_1}\cdot U_{X_2} \cdot U_{X_3}}}$}; 
    \node[anchor=east] at (-2.2cm,0cm) {$\mathbf{H_{P_{X_1X_2X_3}}}$}; 
    \draw[decorate,decoration=brace] (-2.1,-1.3) -- (-2.1,2.9){};
    \draw (-1.2,-1.7) --  (-1.7cm,-2.2cm)  node[anchor=north]
        {$\mathbf{H_{P_{X_1} \cdot P_{X_2} \cdot P_{X_3}}}$}; 
    \draw (0,-1.5) --  (-0.1cm,-2.2cm)  node[anchor=north] {$\mathbf{H_{P_{X_1}}}$}; 
    \draw (1.5,-1.5) --  (1.6cm,-2.2cm)  node[anchor=north west ] {$\mathbf{H_{P_{X_2}}}$}; 
    \draw (1.85,2.5) --  (3.8cm,2.8cm)  node[anchor=west] {$\mathbf{H_{P_{X_3}}}$}; 

  \end{tikzpicture}

%% file: mod_idiagram_color_multivariate.tex
\def\firstcircle{(0,0) circle (1.5cm)}
\def\secondcircle{(0:1.5cm) circle (1.5cm)}
\def\firstellipse{(0,0) cloud (1.5cm)}
\def\secondellipse{(0,1.5cm) cloud (1.5cm)}

\colorlet{circle edge}{black!80}
\colorlet{circle area}{red!60}
\colorlet{ellipse area}{red!60}
\colorlet{ellipse edge}{black!80}

\tikzset{filled/.style={fill=circle area, draw=circle edge, thick},
    outline/.style={draw=circle edge, thick}}
\tikzset{bold/.style={font={\bfseries}}}

\begin{tikzpicture} 
   [thick,scale=0.75,every node/.style={scale=0.75},even odd rule]
    \draw[filled,color=yellow!60,outline=black!80] (-2cm,-2cm) rectangle (3.5cm,2.8cm);
    \draw[filled,color=green!60,outline=black!80] (-1.7cm,-1.7cm) rectangle (3.2cm,2.1cm);
	\node[ 
		cloud, 
		fill=ellipse area,
		minimum width=3cm, 
		minimum height=3cm, 
		align=center, draw] 
		(cloudX) at (0cm, 0cm) {} node[anchor=east] {$\mathbf{H_{P_{\overline X|\overline Y}}}$};
	\node[cloud, 
		fill=ellipse area,
		minimum width=3cm, 
		minimum height=3cm, 
		align=center,  draw] 
		(cloudY) at (1.5cm, 0cm) {} 
		node[anchor=east,bold] at (2.7cm,0) {$\mathbf{H_{P_{\overline Y|\overline X}}}$};

\node[cloud, fill=green!60, 
		minimum width=1.5cm, 
		minimum height=2.5cm, 
		align=center, draw] 
		(intersection) at (0.75cm, 0cm) {};
    \node[anchor=south,bold] at (0.8cm,-0.3) {$\mathbf{I_{P_{\overline X \overline Y}}}$};
    \node[anchor=south] at (0.8cm,1.4cm) {$\mathbf{I_{P_{\overline X \overline Y}}}$};
    \node
         at (0.8cm,2.4cm) {$\Delta\mathbf{H_{P_{\overline X}\cdot P_{\overline Y}}}$};
    \draw (2.5,-2)   --  (2.7cm,-2.2cm) node[anchor=north west]
    {$\mathbf{H_{U_{\overline X}\cdot U_{\overline Y}}}$}; 
   \node[anchor=east] at (-2.2cm,0cm) {$\mathbf{H_{P_{\overline X \overline  Y}}}$}; 
    \draw[decorate,decoration=brace] (-2.1,-1.3) -- (-2.1,1.3){};
    \draw (-1.2,-1.7) --  (-1.7cm,-2.2cm)  node[anchor=north]
        {$\mathbf{H_{P_{\overline X} \cdot P_{\overline Y}}}$}; 
    \draw (0,-1.5) --  (-0.1cm,-2.2cm)  node[anchor=north] {$\mathbf{H_{P_{\overline X}}}$}; 
    \draw (1.5,-1.5) --  (1.6cm,-2.2cm)  node[anchor=north west ] {$\mathbf{H_{P_{\overline Y}}}$}; 
  \end{tikzpicture}

%% file: resultsCMET_R1.tex
Our goal is now to find a decomposition of the entropies around characterizing a joint distribution $P_{\overline X \overline Y}$ between random vectors $\overline X$ and $\overline Y$ in ways analogous to those of \eqref{eq:cbet:balance} 
but considering multivariate input and output. 

Note that it provides no advantage trying to do this on continuous distributions, as the entropic measures used are basic. Rather, what we actually capitalize on is in the outstanding existence of a balance equation between these apparently simple entropic concepts, and what their intuitive meanings afford to the problem of measuring the transfer of information in data processing tasks. 
%
As we set out to demonstrate in this section, our main results are in complete analogy to those of the binary case, but with the flavour of the multivariate case.

\subsection{The Aggregate and Split Channel Multivariate Balance Equation}
\label{sec:theory}
\input{theoryCMET_R1.tex}

\subsection{Visualizations: From i-Diagrams to Entropy Triangles}
\label{sec:visual}
\input{visualizationCMET_R1.tex}

\subsection{Example application: the analysis of feature transformation and selection with entropy triangles}
\label{sec:app}
\input{application_R1.tex}


\subsection{Discussion}
\label{sec:discuss}
\input{discussionCMET_R1.tex}

%% file: theoryCMET_R1.tex
Consider the modified information diagram of Figure~\ref{fig:midiagrams}.\subref{fig:cmet:idiagram} highlighting entropies for some distributions around $P_{\overline X \overline Y}$. 
When we distinguish two random vectors in the set of variables $\overline X$ and $\overline Y$, 
a proper multivariate generalization of the variation of information 
in \eqref{eq:cbet:vi} is
\begin{align}
\label{eq:cmet:vi}
VI_{P_{\overline X\overline Y}} = H_{P_{\overline X| \overline Y}} +  H_{P_{\overline Y| \overline X}}\,.
\end{align}
and we will also call it the \emph{variation of information.}
It represents the addition of the information in $\overline X$ not shared with $\overline Y$ and vice-versa, as captured by the red area in Figure~\ref{fig:midiagrams}.\subref{fig:cmet:idiagram}.
Note that this is a non-negative quantity, since its is the addition of two entropies. 

Next,  consider
\begin{itemize*}[label={}]%
\item $U_{\overline X \overline Y}$\,, the uniform distribution over the supports of $\overline X$ and $\overline Y$,  and 

\item $P_{\overline X}\times{P_{\overline Y}}$\,, the distribution created with the marginals of $P_{\overline X \overline Y}$ considered independent.
\end{itemize*}
Then, we may define a \emph{multivariate divergence with respect to uniformity}---in analogy to \eqref{eq:cbet:delta}---as
\begin{align}
\label{eq:cmet:delta}
\Delta H_{P_{\overline X}\times P_{\overline Y}} = H_{U_{\overline X\overline Y}}- H_{P_{\overline X}\times  P_{\overline Y}}\,.
\end{align}
This is the yellow area in Figure~\ref{fig:midiagrams}.\subref{fig:cmet:idiagram} representing the divergence of the virtual distribution $P_{\overline X}\times P_{\overline Y}$ with respect to uniformity. The virtuality comes from the fact that this distribution does not properly exist in the context being studied. 
Rather, it only appears in the extreme situation that the marginals of $P_{\overline X\overline Y}$ are independent. 

Furthermore, 
recall that both the total entropy of the uniform distribution and the divergence from uniformity factor into individual equalities $H_{U_{\overline X}U_{\overline Y}} = H_{U_{\overline X}} + H_{U_{\overline Y}}$---since uniform joint distributions always have independent marginals---and $H_{P_{\overline X}\times P_{\overline Y}} = H_{P_{\overline X}} + H_{P_{\overline Y}}$. 
Therefore \eqref{eq:cmet:delta} admits splitting as 
$\Delta H_{P_{\overline X}\times P_{\overline Y}} = \Delta H_{P_{\overline X}}  + \Delta H_{P_{\overline Y}} $ 
where 
\begin{align}
\label{eq:cmet:delta:split}
\Delta H_{P_{\overline X}} &= H_{U_{\overline X}}- H_{P_{\overline X}}
&
\Delta H_{P_{\overline Y}} &= H_{U_{\overline Y}}- H_{P_{\overline Y}}\,.
\end{align}
Now, both $U_{\overline X}$ and $U_{\overline Y}$ are the most entropic distributions definable in the support of $\overline X$ and $\overline Y$  whence both $\Delta H_{P_{\overline X}}$ and $\Delta H_{P_{\overline Y}}$ are non-negative, as is their addition.
%
These generalizations are straightforward and intuitively mean that \emph{we expect them to agree with the intuitions developed in the CBET}, which is an important usability concern. 

The problem is finding a quantity that fulfills the same role as the (bivariate) mutual information. 
The first property that we would like to have is for this quantity to be a ``transmitted information'' after conditioning away any of the entropy of either partition, so we propose the following as a definition: 
\begin{align}
\label{eq:bi:internal}
I_{P_{\overline X \overline Y}} = H_{P_{\overline X \overline Y}} - VI_{P_{\overline X \overline Y}}
\end{align}
represented by the inner green area in the i-diagram of Figure~\ref{fig:midiagrams}.\subref{fig:cmet:idiagram}. 
%
This can easily be ``refocused'' on each of the subsets of the partition: 
\begin{Lemma}
Let $P_{\overline X \overline Y}$  be a discrete joint distribution. Then
\begin{align}
\label{eq:bi:def}
	H_{P_{\overline X}} - H_{P_{\overline X \mid \overline Y}}
	= H_{P_{\overline Y}} - H_{P_{\overline Y \mid \overline X}}
	= I_{P_{\overline X \overline Y}} 
\end{align}
\end{Lemma}
\begin{proof}
Recalling that the conditional entropies are easily related to the joint entropy by the chain rule 
$H_{P_{\overline X \overline Y}} = H_{P_{\overline X}} + H_{P_{\overline Y \mid \overline X}}
	= H_{P_{\overline Y}} + H_{P_{\overline X \mid \overline Y}}$, 
simply subtract $VI_{P_{\overline X \overline Y}}$.
\end{proof}
This property introduces the notion that this information is \emph{within} each of $\overline X$ and $\overline Y$ \emph{independently but mutually induced.}
It is easy to see that this quantity appears once again in the i-diagram: 
\begin{Lemma}
Let $P_{\overline X \overline Y}$  be a discrete joint distribution. Then
\begin{align}
\label{eq:bi:external}
I_{P_{\overline X \overline Y}} =  H_{P_{\overline X}\times{P_{\overline Y}}}  - H_{P_{\overline X \overline Y}}\,. 
\end{align}
\end{Lemma}
\begin{proof}
Considering the entropy decomposition of $P_{\overline X}\times{P_{\overline Y}}$:
\begin{align*}
H_{P_{\overline X}\times{P_{\overline Y}}}  - H_{P_{\overline X \overline Y}} 
	=  H_{P_{\overline X}} + H_{P_{\overline Y}}  
				- \left( H_{P_{\overline Y}} + H_{P_{\overline X \mid \overline Y}} \right) 
	= H_{P_{\overline X}} - H_{P_{\overline X \mid \overline Y}} 
	= I_{P_{\overline X \overline Y}}
\end{align*}
\end{proof}
In other words, this is the quantity of information required to bind $P_{\overline X}$ and $P_{\overline Y}$; equivalently,  it is the amount of information \emph{lost} from $P_{\overline X} \times P_{\overline Y}$ to achieve the binding in $P_{\overline X \overline Y}$.
Pictorially, this is the outermost green area in Fig.~\ref{fig:midiagrams}.\subref{fig:cmet:idiagram}, and \emph{it must be non-negative}, since $P_{\overline X} \times P_{\overline Y}$ is more entropic than $P_{\overline X \overline Y}$. 
Notice that \eqref{eq:bi:internal} and \eqref{eq:bi:def}
are the analogues of \eqref{def:tc} and \eqref{def:dtc},  
respectively, but with the flavor of \eqref{eq:cbet:mi} and \eqref{eq:cbet:mi:alt}.
Therefore, this quantity must be the multivariate mutual information of $P_{\overline X \overline Y}$ as per the Kullback-Leibler divergence definition:
\begin{Lemma}
Let $P_{\overline X \overline Y}$  be a discrete joint distribution. Then
\begin{align}
\label{eq:bi:div}
I_{P_{\overline X \overline Y}} = \sum_{i,j} P_{\overline X \overline Y}(x_i,y_j) \log\frac{P_{\overline X \overline Y}(x_i,y_j)}{P_{\overline X }(x_i)P_{\overline Y}(y_j)}
\end{align}
\end{Lemma}
\begin{proof}
This is an easy manipulation. 
\begin{align*}
\sum_{i,j} P_{\overline X \overline Y}(x_i,y_j) \log\frac{P_{\overline X \overline Y}(x_i,y_j)}{P_{\overline X }(x_i)P_{\overline Y}(y_j)} &=
\sum_{i,j} P_{\overline X \overline Y}(x_i,y_j) \log\frac{P_{\overline X \mid \overline Y= y_j}(x_i |y_j)}{P_{\overline X }(x_i)} = 
\sum_{i} P_{\overline X }(x_i) \log\frac{1}{P_{\overline X }(x_i)} - 
\\
&- 
\sum_j P_{\overline Y}(y_j) \sum_{i} P_{\overline X \mid \overline Y= y_j}(x_i |y_j) \log\frac{1}{P_{\overline X \mid \overline Y= y_j}(x_i |y_j)} = 
\\
= H_{P_{\overline X}} - H_{P_{\overline X\mid \overline Y}} = I_{P_{\overline X \overline Y}}, 
\end{align*}
after a step of marginalization and considering \eqref{eq:cbet:mi:alt}. 
\end{proof}

With these relations we can state our first theorem:
\begin{Theorem}
\label{theo:cmbe}
Let $P_{\overline X \overline Y}$  be a discrete joint distribution.
Then the following decomposition holds: 
\begin{align}
  \label{eq:cmet:balance}
  H_{U_{\overline X}\times U_{\overline Y}} &= \Delta H_{P_{\overline X}\times P_{\overline Y}} +  2* I_{P_{\overline X \overline Y}} +  VI_{P_{\overline X \overline Y}} 
  \\
  0 &\leq \Delta H_{P_{\overline X}\times P_{\overline Y}}, I_{P_{\overline X \overline Y}},  VI_{P_{\overline X \overline Y}} \leq H_{U_{\overline X}\times U_{\overline Y}} \notag
\end{align}
\end{Theorem}
\begin{proof}
From \eqref{eq:cmet:delta} we have 
$H_{U_{\overline X}\times U_{\overline Y}} = \Delta H_{P_{\overline X}\times P_{\overline Y}} + H_{P_{\overline X}\times P_{\overline Y}}$ whence by introducing \eqref{eq:bi:internal} and \eqref{eq:bi:external} we obtain:
\begin{align}
\label{eq:cmet:be}
H_{U_{\overline X}\times U_{\overline Y}} = 
	\Delta H_{P_{\overline X}\times P_{\overline Y}} + 
	I_{P_{\overline X \overline Y}} +  
	H_{P_{\overline X\overline Y}} = 
	\Delta H_{P_{\overline X}\times P_{\overline Y}} + 
	I_{P_{\overline X \overline Y}} +  
	I_{P_{\overline X \overline Y}} +  
	VI_{P_{\overline X\overline Y}}.
\end{align}
Recall that each quantity is non-negative by \eqref{eq:cmet:vi},  \eqref{eq:cmet:delta}  and \eqref{eq:bi:div}, 
so the only things left to be proven are the limits for each quantity in the decomposition. 
For that purpose, consider the following clarifying \emph{conditions},
\begin{enumerate}
\item \textbf{$\overline{X}$ marginal uniformity} when $H_{P_{\overline X}} = H_{U_{\overline X}}$, 
\textbf{$\overline{Y}$ marginal uniformity} when $H_{P_{\overline Y}} = H_{U_{\overline Y}}$ 
and \textbf{marginal uniformity} when both conditions coocur. 

\item \textbf{Marginal independence}, when $P_{\overline X\overline Y} = P_{\overline X}\times P_{\overline Y}$.

\item \textbf{$\overline Y$ determines $\overline X$} when $H_{P_{\overline X \mid \overline Y}}=0$,
\textbf{$\overline X$ determines $\overline Y$} when $H_{P_{\overline Y \mid \overline X}}=0$
and \textbf{mutual determination}, when both conditions hold. 
\end{enumerate}
Notice that these conditions are \emph{independent of each other} and that \emph{each fixex the value of one of the quantities in the balance}: 
\begin{itemize}
\item for instance, in case $H_{P_{\overline X}} = H_{U_{\overline X}}$ then $\Delta H_{P_{\overline X}} = 0$ after \eqref{eq:cmet:delta:split}. Similarly,  if $H_{P_{\overline Y}} = H_{U_{\overline Y}}$ then $\Delta H_{P_{\overline Y}} = 0$. Hence when marginal uniformity holds, we have $\Delta H_{P_{\overline X\overline Y}} = 0$. 
\item Similarly, when marginal independence holds, we see that $I_{P_{\overline X \mid \overline Y}} =0$ from \eqref{eq:bi:external}.  
Otherwise stated, $H_{P_{\overline X \mid \overline Y}}  = H_{P_{\overline X}}$ and $H_{P_{\overline Y \mid \overline X}}  = H_{P_{\overline Y}}$\,. 
\item Finally, if mutual determination holds---that is to say the variables in either set are deterministic functions of those of the other set---by the definition of the multivariate variation of information, we have $VI_{P_{\overline X \mid \overline Y}}=0$.  

\end{itemize}
Therefore, these three conditions fix the lower bounds for their respectively related quantities. 
Likewise, the upper bounds hold when \emph{two} of the conditions hold at the same time. This is easily seen invoking the previously found balance equation \eqref{eq:cmet:be}:
\begin{itemize}
\item For instance, if marginal uniformity  holds, then $\Delta H_{P_{\overline X\overline Y}} = 0$\,. But if marginal independence also holds, then $I_{P_{\overline X \mid \overline Y}} =0$ whence by \eqref{eq:cmet:be} $VI_{P_{\overline X\overline Y}} = H_{U_{\overline X}\times U_{\overline Y}}$. 

\item But if both marginal uniformity and mutual determination hold, then we have $\Delta H_{P_{\overline X\overline Y}} = 0$ and $VI_{P_{\overline X\overline Y}}=0$  so that $I_{P_{\overline X \overline Y}}  = H_{U_{\overline X}\times U_{\overline Y}}$.

\item Finally, if both mutual determination and marginal indepence holds, then a fortiori $\Delta H_{P_{\overline X \overline Y}}  = H_{U_{\overline X}\times U_{\overline Y}}$.
\end{itemize}
This concludes the proof. 
\end{proof}
Notice how the bounds also allow an interpretation similar to that of \eqref{eq:cbet:balance}. In particular, the interpretation of the conditions for actual joint distributions will be taken again in Section~\ref{sec:visual}. 

The next question is whether the balance equation also admits splitting. 
\begin{Theorem}
Let $P_{\overline X \overline Y}$  be a discrete joint distribution. 
Then the Channel Multivariate Entropy Balance equation can be split as:
\begin{align}
\label{eq:cmet:split:X}
  H_{U_{\overline X}} &= \Delta H_{P_{\overline X}} +  I_{P_{\overline X \overline Y}} +  H_{P_{\overline X\mid \overline Y}} 
  &
  0 &\leq \Delta H_{P_{\overline X}}, I_{P_{\overline X \overline Y}},  H_{P_{\overline X |\overline Y}} \leq H_{U_{\overline X}}
\\
\label{eq:cmet:split:Y}
  H_{U_{\overline Y}} &= \Delta H_{P_{\overline Y}} +  I_{P_{\overline X \overline Y}} +  H_{P_{\overline Y \mid \overline X}}
   &
   0 &\leq \Delta H_{P_{\overline Y}}, I_{P_{\overline X \overline Y}},  H_{P_{\overline Y |\overline X}} \leq H_{U_{\overline Y}}
\end{align}
\end{Theorem}
\begin{proof}
We prove \eqref{eq:cmet:split:X}: the proof of \eqref{eq:cmet:split:Y} is similar \emph{mutatis mutandis.}

In a similar way as for~\eqref{eq:cmet:balance}, we have that $H_{U_{\overline X}} = \Delta H_{P_{\overline X}} + H_{P_{\overline X}}$. By introducing the 
value of  $ H_{P_{\overline X}}$
from \eqref{eq:bi:def} we obtain the decomposition of $H_{U_{\overline X}}$ of \eqref{eq:cmet:split:X}. 

These quantities are non-negative, as mentioned. 
Next consider the $\overline X$ marginal uniformity condition applied to the input vector introduced in the proof of Theorem~\ref{theo:cmbe}. Clearly,  $\Delta H_{\overline X}=0$. 
Marginal independence, again, is the condition so that $I_{\overline X \overline Y} = 0$. 
Finally, if $\overline Y$ determines $\overline X$ then $H_{P_{\overline X\mid \overline Y}}=0$. 
These conditions individually provide the lower bounds on each quantity. 


On the other hand, when we put together any two of these conditions, we obtain the upper bound for the unspecified variable:
so, if $\Delta H_{P_{\overline X}} = 0$ and $I_{P_{\overline X \overline Y}}=0$ then $H_{P_{\overline X \mid \overline Y}} = H_{P_{\overline X}} = H_{U_{\overline X}}$.
Also, if $I_{P_{\overline X \overline Y}}=0$ and $H_{P_{\overline X \mid \overline Y}}=0$, then $H_{P_{\overline X }}=H_{P_{\overline X \mid \overline Y}}=0$ and $\Delta H_{P_{\overline X}} = H_{U_{\overline X}} - 0$\,.  
Finally, if $H_{P_{\overline X \mid \overline Y}}=0$ and $\Delta H_{P_{\overline X}} = 0$, then $I_{P_{\overline X \overline Y}} = H_{P_{\overline X}} - H_{P_{\overline X  \mid \overline Y}} = H_{U_{\overline X}} - 0$\,. 
\end{proof}

%% file: visualizationCMET_R1.tex
\subsubsection{The Channel Multivariate Entropy triangle}
Our next goal is to develop an exploratory analysis tool similar to the CBET introduced in Section~\ref{sec:et}.
As in that case, 
we need the equation of a simplex to represent the information balance of a multivariate transformation. For that purpose, as in   \eqref{eq:cbet:simplex}
%
we may normalize  by the overall entropy $H_{U_{\overline X}\times U_{\overline Y}} $ to obtain the  equation of the 2-simplex in multivariate entropic space,
\begin{align}
\label{eq:cmet:simplex}
  1 &= \Delta'H_{P_{\overline X}\times P_{\overline Y}} +  2* I'_{P_{\overline X \overline Y}} +  VI'_{P_{\overline X \overline Y}} 
\\ 
 0 &\leq \Delta'H_{P_{\overline X}\times P_{\overline Y}}, I'_{P_{\overline X \overline Y}}, VI'_{P_{\overline X \overline Y}} \leq  1\,.\notag
\end{align}
The de Finetti diagram of this equation then provides the aggregated \emph{Channel Multivariate Entropy Triangle, CMET}. 

A \emph{formal} graphical assessment of multivariate joint distribution with the CMET is fairly simple using the schematic in
Fig.~\ref{fig:schema:CMET:formal}.\subref{fig:schema:CMET:formal:agg} and the conditions of Theorem~\ref{theo:cmbe}:
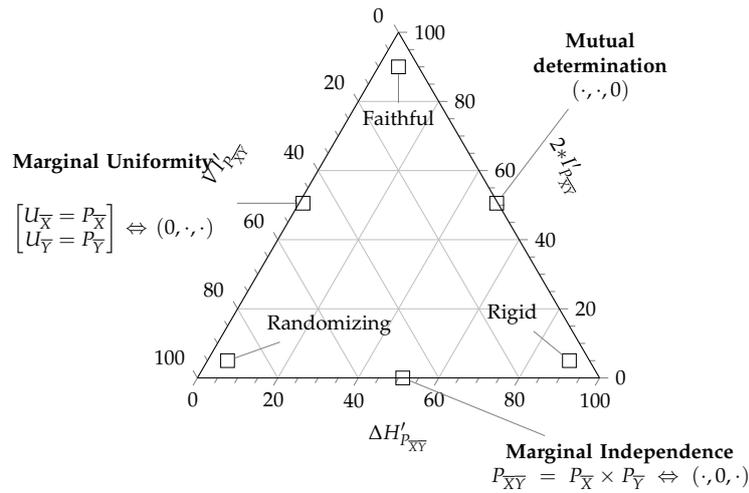
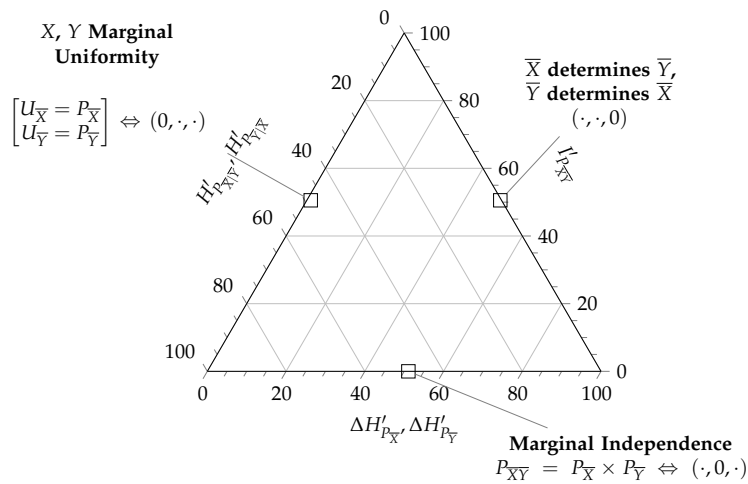
\begin{figure*}[!tb]
 \centering
 \subfloat[Schematic CMET with a formal interpretation.]{
 	\resizebox{0.7\linewidth}{!}{
  		\input{annotated_triangle_CMET_agg_formal.tex}
  		}
  \label{fig:schema:CMET:formal:agg}  		
  	}
  	\\
  \subfloat[Schematic \emph{split CMETs} with formal interpretations. Note that there are \textbf{two types} of overimposed entropy triangles in this figure.]{
 	\resizebox{0.7\linewidth}{!}{
  		\input{annotated_triangle_CMET_split_formal.tex}
  		}
  \label{fig:schema:CMET:formal:split}
  	}
  \caption{ 
  \textbf{Schematic Channel Multivariate Entropy Triangles (CMET) showing interpretable zones and extreme cases using formal conditions.} The annotations on the center of each side are meant to hold for that whole side, those for the vertices are meant to hold in their immediate neighborhood too.}
  \label{fig:schema:CMET:formal}
\end{figure*}
\begin{itemize}
\item The lower side of the triangle with $I'_{P_{\overline X \overline Y}}=0$, affected of \emph{marginal independence} $P_{\overline X \overline Y} = P_{\overline X}\times P_{\overline Y}$, is the locus of partitioned joint distributions who do not share information between the two blocks $\overline X$ and $\overline Y$. 

\item The right side of the triangle with $VI'_{P_{\overline X \overline Y}}=0$, described with \emph{mutual determination} $H'_{P_{\overline X \mid \overline Y}}=0=H'_{P_{\overline Y \mid \overline X}}$, is the locus of partitioned joint distributions whose groups do not carry supplementary information to that provided by the other group.

\item The left sidewith $\Delta H'_{P_{\overline X \overline Y}}=0$, describing distributions with \emph{uniform marginals} $P_{\overline X} = U_{\overline X}$ and $P_{\overline Y} = U_{\overline Y}$, is the locus of partitioned joint distributions that offer as much potential information for transformations as possible.
\end{itemize}
Based on these characterizations we can attach interpretations to other regions of the CMET:
\begin{itemize}
\item If we want a transformation from $\overline X$ to $\overline Y$ to be \emph{faithful}, then we want to maximize the information used for mutual determination $I'_{P_{\overline X \overline Y}}\to 1$, equivalently, minimize at the same time the divergence from uniformity $\Delta H'_{P_{\overline X \overline Y}} \to 0$ and the information that only pertains to each of the blocks in the partition $VI'_{P_{\overline X \overline Y}} \to 0$. So the coordinates of a faithful partitioned joint distribution will lay close to the apex of the triangle.

\item However, if the coordinates of a distribution lay close to the left vertex $VI'_{P_{\overline X \overline Y}} \to 1$, then it shows marginal uniformity $\Delta H'_{P_{\overline X \overline Y}} \to 0$ but shares little or no information between the blocks $I'_{P_{\overline X \overline Y}}\to 0$, hence it must be a \emph{randomizing} transformation.

\item Distributions whose coordinates lay close to the right vertex $\Delta H'_{P_{\overline X \overline Y}} \to 1$ are essentially deterministic and in that sense carry no information $I'_{P_{\overline X \overline Y}}\to 0, VI'_{P_{\overline X \overline Y}} \to 0$. Indeed in this instance there does not seem to exist a transformation, whence we call them \emph{rigid}. 

\end{itemize}
These qualities are annotated on the vertices of the schematic CMET of Fig.~\ref{fig:schema:CMET:formal}.\subref{fig:schema:CMET:formal:agg}. 
Note that different applications may call for partitioned distributions with different qualities and the one used above is pertinent when the partitioned joint distributions models a transformation of $\overline X$ into $\overline Y$ or vice-versa.

\subsubsection{Normalized Split Channel Multivariate Balance Equations}
With a normalization similar to that from  \eqref{eq:cbet:split:balance} to \eqref{eq:cbet:split:simplex},  \eqref{eq:cmet:split:X} and~\eqref{eq:cmet:split:Y}
 naturally lead  to 2-simplex equations normalizing by $H_{U_{\overline X}}$ and $H_{U_{\overline Y}}$, respectively
\begin{align}
\label{eq:cmet:split:simplex:X}
  1 &= \Delta'H_{P_{\overline X}} +  I'_{P_{\overline X \overline Y}} +  H'_{P_{\overline X\mid \overline Y}} 
\\ 
  0 &\leq \Delta'H_{P_{\overline X}}, I'_{P_{\overline X \overline Y}},  H'_{P_{\overline X |\overline Y}} \leq 1 \notag
\\
\label{eq:cmet:split:simplex:Y}
  1 &= \Delta'H_{P_{\overline Y}} +  I'_{P_{\overline X \overline Y}} +  H'_{P_{\overline Y \mid \overline X}}
\\ 
   0 &\leq \Delta'H_{P_{\overline Y}}, I'_{P_{\overline X \overline Y}},  H'_{P_{\overline Y |\overline X}} \leq 1 \notag
\end{align}
Note that the quantities $\Delta H'_{P_{\overline X}}$ and $\Delta H'_{P_{\overline Y}}$ have been independently motivated and named \emph{redundancies}~\citep[\S~2.4]{kay:03}.

These are actually two different representations for each of the two blocks in the partitioned joint distribution. Using the fact that they share one coordinate---$ I'_{P_{\overline X \overline Y}}$---and the rest are analogues---$\Delta'H_{P_{\overline X}}$ and $\Delta'H_{P_{\overline Y}}$ on one side, and $H'_{P_{\overline X\mid \overline Y}} $ and $H'_{P_{\overline Y \mid \overline X}}$ on the other---we can represent both equations \emph{at the same time} in a single de Finetti diagram. We call this representation the \emph{split Channel Multivariate Entropy Triangle}, an schema of which can be seen in Fig.~\ref{fig:schema:CMET:formal}.\subref{fig:schema:CMET:formal:split}. 
The qualifying ``split'' then refers to the fact that each partitioned joint distribution appears as \emph{two points} in the diagram. 
Note the double annotation in the left and bottom coordinates implying that there are \emph{two} different diagrams overlapping. 

Conventionally, the point referring to the $\overline X$ block described by \eqref{eq:cmet:split:simplex:X} is represented with a cross, while the point referring to the $\overline Y$ block described by \eqref{eq:cmet:split:simplex:Y} is represented with a circle as will be noted in Figure \ref{fig:transfo:independent}.

The formal interpretation of this split diagram with the conditions of Theorem~\ref{theo:cmbe} follows that of the aggregated CMET but considering only one block at a time, for instance, for $\overline X$: 
\begin{itemize}
\item The lower side of the triangle is interpreted as before.

\item The right side of the triangle is the locus of the partitioned joint distribution whose $\overline X$ block is completely determined by the $\overline Y$ block, that is, $H'_{P_{\overline X\mid \overline Y}} =0$.

\item The left side of the triangle $\Delta H'_{P_{\overline X}} =0$ is the locus of those partitioned joint distributions whose $\overline X$ marginal is uniform $P_{\overline X} = U_{\overline X}$\,. 

\end{itemize}
The interpretation is analogue for $\overline Y$ \emph{mutatis mutandis}.

The purpose of this representation is to investigate the formal conditions separately on each block. However, for this split representation we have to take into consideration that the normalizations may not be the same, that is $H_{P_{\overline X}}$ and $H_{P_{\overline Y}}$ are, in general, different.


A full example of the interpretation of both types of diagrams, the CMET and the split CMET is provided in the next Section in the context of feature transformation and selection.

%% file: annotated_triangle_CMET_agg_formal.tex

\begin{tikzpicture} 

\begin{ternaryaxis}[
 xmin=0,
 xmax=100,
 ymin=0,
 ymax=100,
 zmin=0,
 zmax=100, 
 xlabel=$2{\ast}I'_{P_{\overline X \overline Y}}$,
 ylabel=$VI'_{P_{\overline X \overline Y}}$,
 zlabel=$\Delta H'_{P_{\overline X \overline Y}}$,
 label style={sloped},
 minor tick num=3,
 grid=major
]

\node[pin=270:Faithful,draw=black] at (axis cs:90,5) {};
\node[pin=130:Rigid,draw=black] at (axis cs:5,5) {};
\node[pin=30:Randomizing,draw=black] at (axis cs:5,90) {};

\end{ternaryaxis}

\node[pin={[pin distance=1cm,text width=6cm,align=center] 
  300:\textbf{Marginal Independence $P_{\overline X \overline Y} = P_{\overline X}\times P_{\overline Y} \Leftrightarrow (\cdot,0,\cdot)$ }}, draw=black]  
   at (3.5,0) {};P
\node[pin={[pin distance=1.5cm,text width=3cm,align=center] 85:\textbf{Mutual determination $  (\cdot,\cdot,0)$ }}, draw=black] at (5.1,3) {};
\node[pin={
	[pin distance=1cm,text width=4cm,align=center] 180:%
		\textbf{Marginal Uniformity\\}\vspace{0.5cm}
$\begin{bmatrix}
      U_{\overline X} = P_{\overline X} \\ U_{\overline Y} = P_{\overline Y}      
    \end{bmatrix}
  \Leftrightarrow (0,\cdot,\cdot)$
	}, draw=black] at (1.8,3) {};

\end{tikzpicture}







%% file: annotated_triangle_CMET_split_formal.tex

\begin{tikzpicture} 

\begin{ternaryaxis}[
 xmin=0,
 xmax=100,
 ymin=0,
 ymax=100,
 zmin=0,
 zmax=100, 
 xlabel=$I'_{P_{\overline X \overline Y}}$,
 ylabel= 
 {$H'_{P_{\overline X\mid \overline Y}}, H'_{P_{\overline Y \mid \overline X}}$},
 zlabel={$\Delta H'_{P_{\overline X}}, \Delta H'_{P_{\overline Y}}$},
 label style={sloped},
 minor tick num=3,
 grid=major
]
\end{ternaryaxis}

\node[pin={[pin distance=1cm,text width=6cm,align=center] 
  300:\textbf{Marginal Independence $P_{\overline X \overline Y} = P_{\overline X}\times P_{\overline Y} \Leftrightarrow (\cdot,0,\cdot)$ }}, draw=black]  
   at (3.5,0) {};P
\node[pin={[pin distance=1.0cm,text width=3cm,align=center] 85:%
	\textbf{$\overline X$ determines $\overline Y$, $\overline Y$ determines $\overline X$ \\$  (\cdot,\cdot,0)$ }}, 
	draw=black] at (5.1,3) {};
\node[pin={
	[pin distance=1.5cm,text width=4cm,align=center] 150:%
		\textbf{$X$, $Y$ Marginal Uniformity\\}\vspace{0.5cm}
$\begin{bmatrix}
      U_{\overline X} = P_{\overline X} \\ U_{\overline Y} = P_{\overline Y}      
    \end{bmatrix}
  \Leftrightarrow (0,\cdot,\cdot)$
	}, draw=black] at (1.8,3) {};

\end{tikzpicture}







%% file: application_R1.tex
In this Section we present an application of the results obtained above to a machine learning subtask: the transformation and selection of features for supervised classification.

{%


\textbf{\emph{The task. }}%
An extended practice in supervised classification is to explore different transformations of the observations and then evaluate such different approaches on different classifiers for a particular task~\cite{wit:eib:hal:11}.
Instead of this ``in the loop'' evaluation---that conflates the evaluation of the transformation and the classification---we will use the CMET to evaluate \emph{only} the transformation block using the information transferred from the original to the transformed features as heuristic. 
As specific instances of transformations, we will evaluate the use of Principal Component Analysis (PCA)~\cite{pea:01} and Independent Component Analysis (ICA)~\cite{bel:sej:95} which are often employed for dimensionality reduction. 

Note that we may evaluate feature transformation and dimensionality reduction  at the same time with the techniques developed above: the transformation procedure in the case of PCA and ICA may provide the $\overline Y$ as a ranking of features, so that we may  carry out  \emph{feature selection} afterwards by selecting subsets $\overline Y_i$ spanning from the first-ranked to the $i$-th feature.

\textbf{\emph{The tools. }}%
PCA is a staple technique in statistical data analysis and machine learning based in the Singular Value Decomposition of the data matrix  to obtain projections along the singular vectors that account for its variance in decreasing amount, so PCA ranks the transformed features by this order. 
The implementation used in our examples are those of the publicly available  R packages \texttt{stats} (v. 3.3.3)%
\footnote{\url{https://stat.ethz.ch/R-manual/R-devel/library/stats/html/00Index.html}. Last checked 11/06/2018.}.

While PCA aims at the orthogonalization of the projections, ICA finds the projections, also known as \emph{factors},  by maximimizing their statistical independence, in our example by minimizing  a cost term related to their mutual information~\cite{hyv:oja:00}. 
However, this does not result in a ranking of the transformed features, hence we have created a pseudo-ranking by carrying an ICA transformation  obtaining $i$ transformed features for all sensible values of $1 \leq i \leq n$ using independent runs of the ICA algorithm.  
The implementation used in our examples is that of fastICA~\cite{hyv:oja:00}  as implemented in the R package \texttt{fastICA} (v. 1.2-1)%
\footnote{\url{https://cran.r-project.org/package=fastICA}. Last checked 11/06/2018.} with standard parameter values\footnote{ \texttt{alg.typ}=``parallel'', \texttt{fun}=``logcosh'', \texttt{alpha}=1, \texttt{method}=``C'', \emph{row.norm}= FALSE, \texttt{maxit}=200, \texttt{tol}=0.0001.}.

The entropy diagrams and calculations were carried out with the open-source \texttt{entropies} experimental R package that provides an implementation of the present framework 
\footnote{Available at \url{https://github.com/FJValverde/entropies.git}. Last checked: 11/06/2018.}. 
The analysis carried out in this section is part of an illustrative vignette for the package and will remain so in future releases. 

\textbf{\emph{Analysis of results. }} %
We analized in this way some UCI classification datasets~\cite{bac:lic:13}, whose number of features $n$, classes $K$ and feature vectors $m$ can be seen in Table~\ref{tab:datasets}. 
%
\begin{table}[!th]
\centering
\caption{Datasets analyzed. 
\label{tab:datasets}
}
\begin{tabular}{rlrrr}
  \hline
 & name & K & n & m \\ 
  \hline
  1 & Ionosphere &   2 &  34 & 351 \\ 
  2 & \textbf{Iris} &   3 &   4 & 150 \\ 
  3 & \textbf{Glass} &   7 &   9 & 214 \\ 
  4 & \textbf{Arthritis} &   3 &   3 &  84 \\ 
  5 & BreastCancer &   2 &   9 & 699 \\ 
  6 & Sonar &   2 &  60 & 208 \\ 
  7 & Wine &   3 &  13 & 178 \\ 
   \hline
\end{tabular}
\end{table}
For simplicity issues, we decided to illustrate our new techniques on  three datasets: \emph{Iris}, \emph{Glass} and \emph{Arthritis}. 
\emph{Ionosphere}, \emph{BreastCancer}, \emph{Sonar} and \emph{Wine} have a similar  pattern to \emph{Glass}, but less interesting, as commented below. 
Besides, both \emph{Ionosphere} and \emph{Wine} have too many features for the kind of neat visualization we are trying to use in this paper. 
We have also used a slightly modified entropy triangles in which the colors of the axes are related to those of the information diagrams of Figure~\ref{fig:midiagrams}.\subref{fig:cmet:idiagram} . 
} 

\begin{figure}
	\centering	
	\subfloat[PCA on \emph{Iris}]{
		\includegraphics[width=0.45\linewidth]{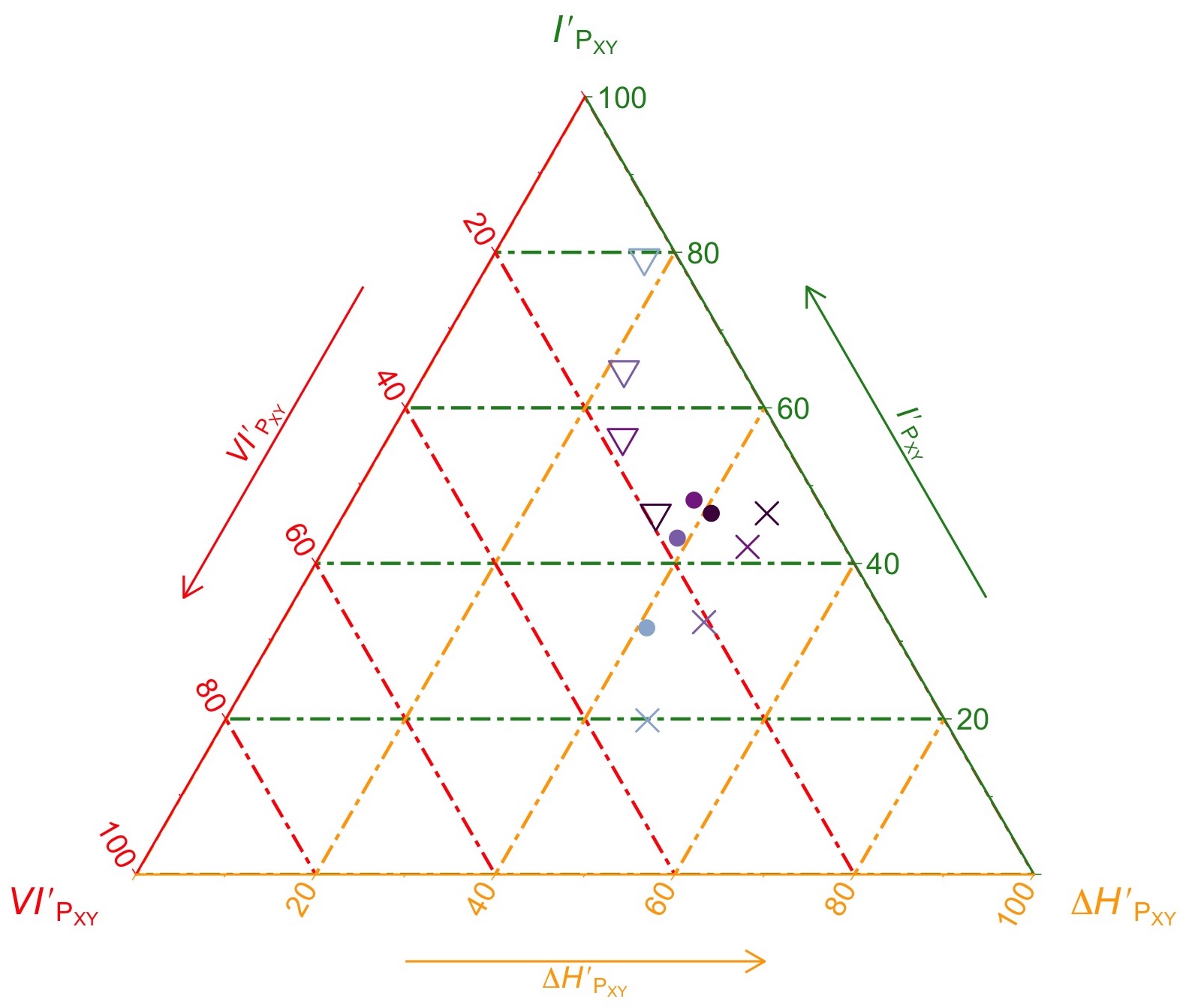}
		\label{fig:transfo:PCA:iris}
	}
	\subfloat[ICA on \protect\emph{Iris}]{
		\includegraphics[width=0.55\linewidth]{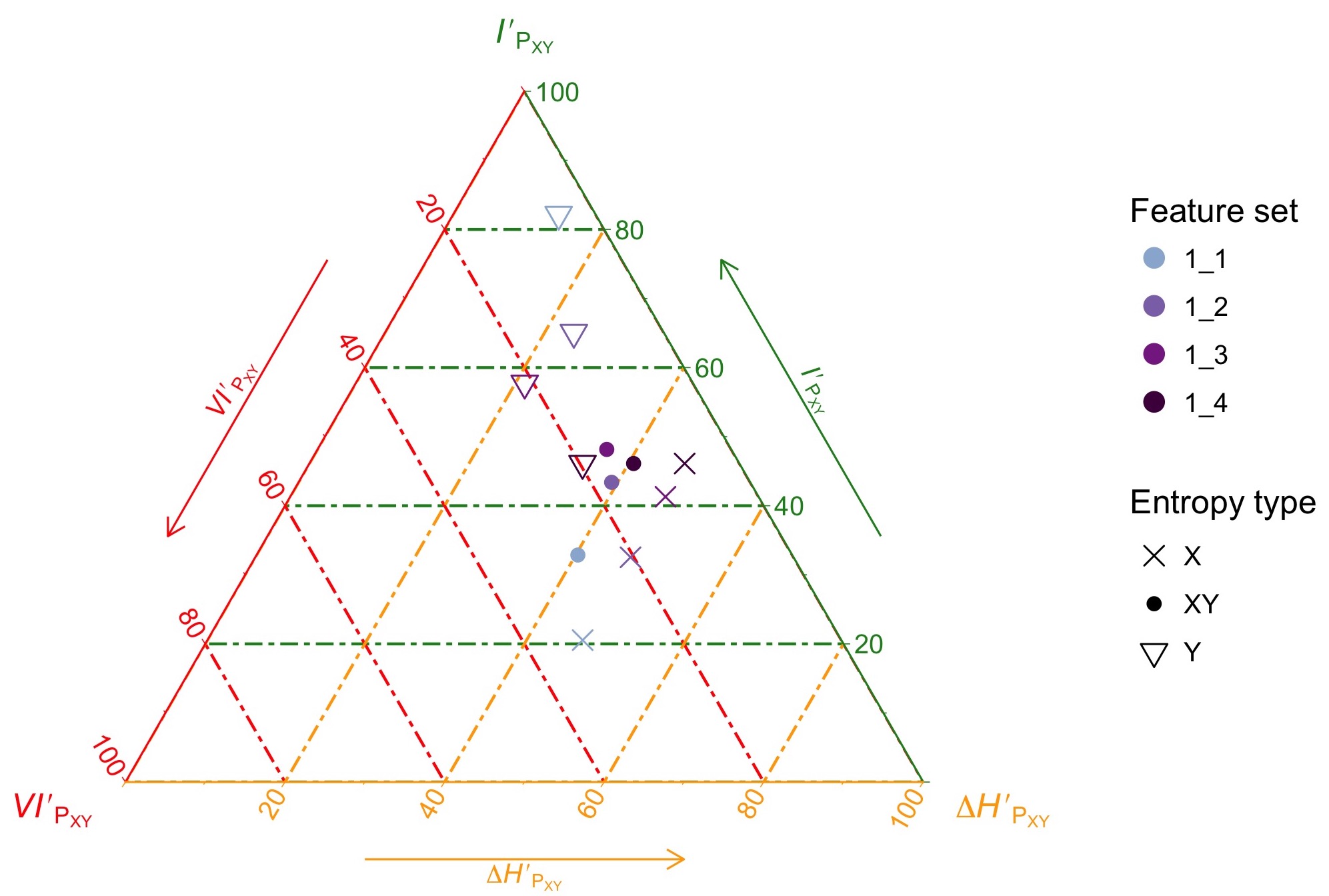}
		\label{fig:transfo:ICA:iris}
	}
	\\
	\subfloat[PCA on \emph{Glass}]{
		\includegraphics[width=0.45\linewidth]{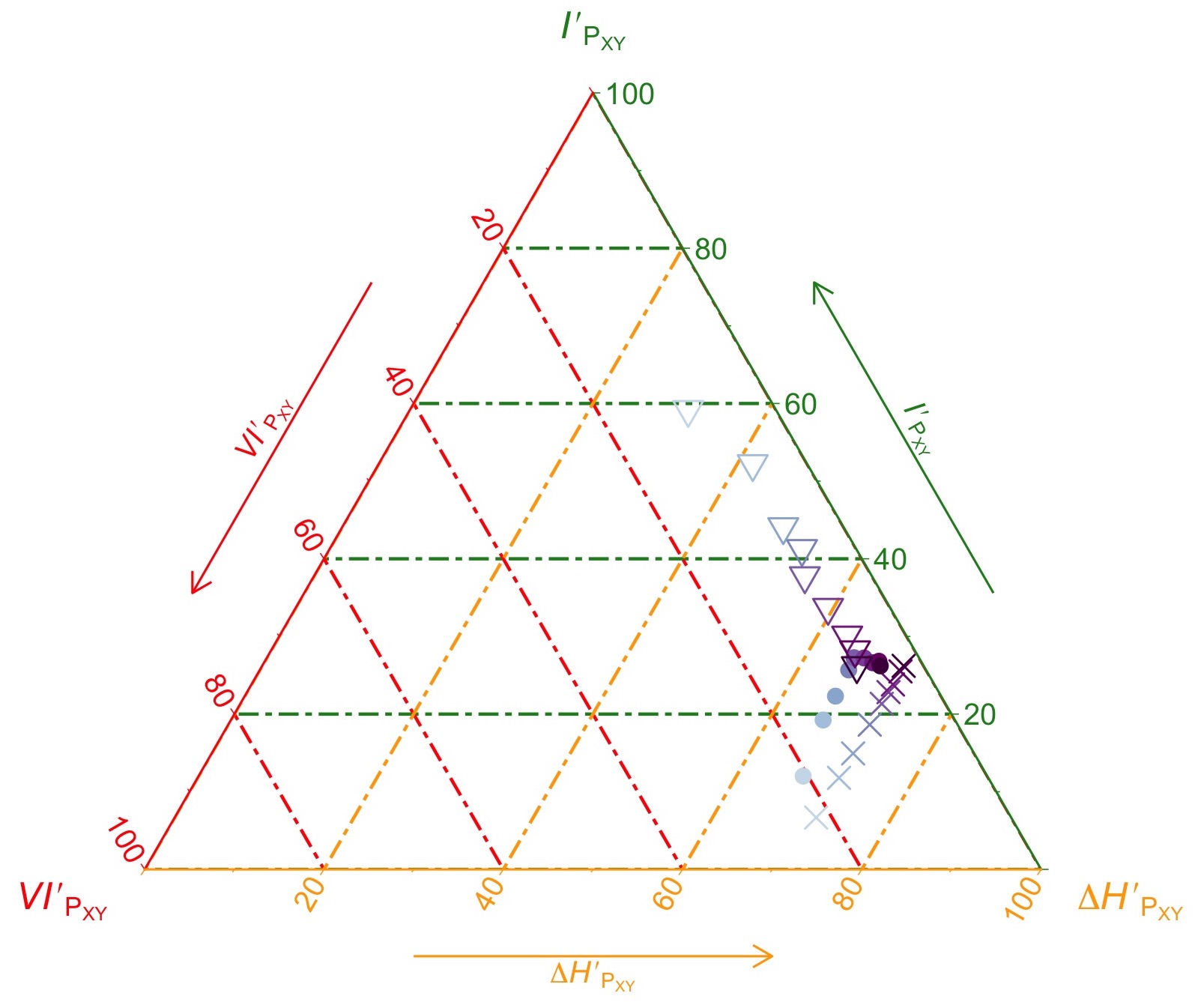}
		\label{fig:transfo:PCA:glass}
	}
	\subfloat[ICA on \protect\emph{Glass}]{
		\includegraphics[width=0.55\linewidth]{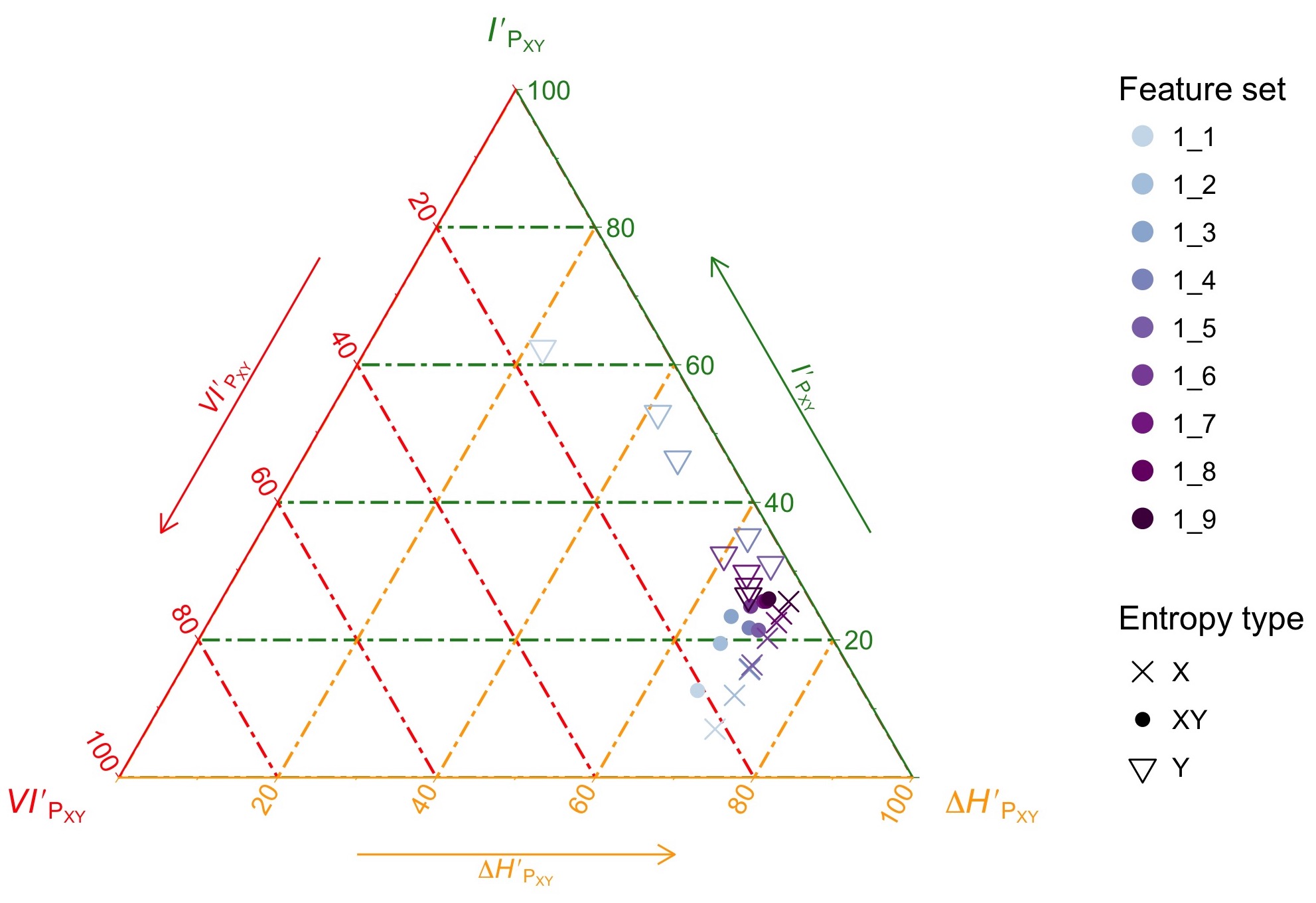}
		\label{fig:transfo:ICA:glass}
	}
	\\
	\subfloat[PCA on \emph{Arthritis}]{
		\includegraphics[width=0.45\linewidth]{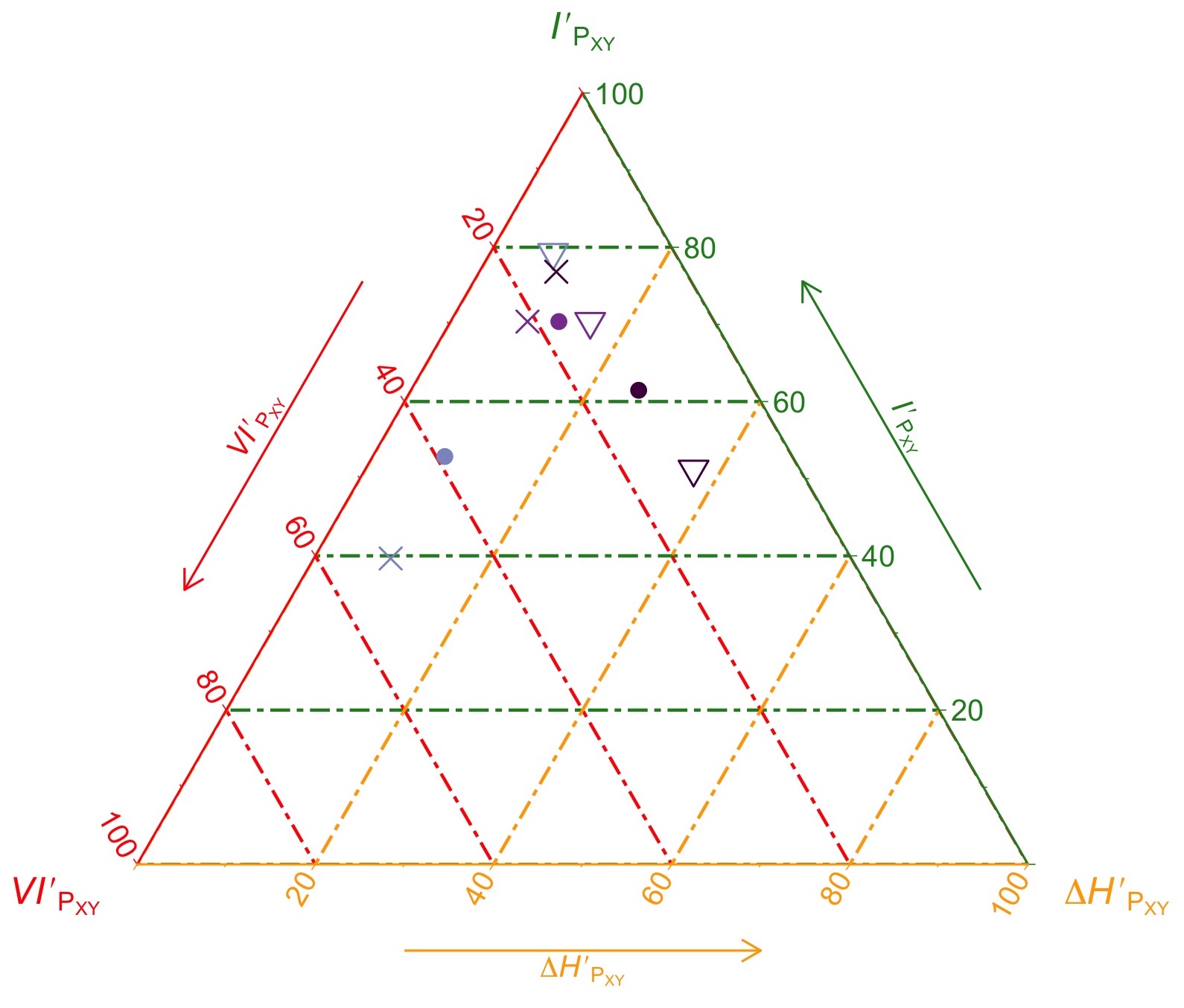}
		\label{fig:transfo:PCA:arthr}
	}
	\subfloat[ICA on \protect\emph{Arthritis}]{
		\includegraphics[width=0.55\linewidth]{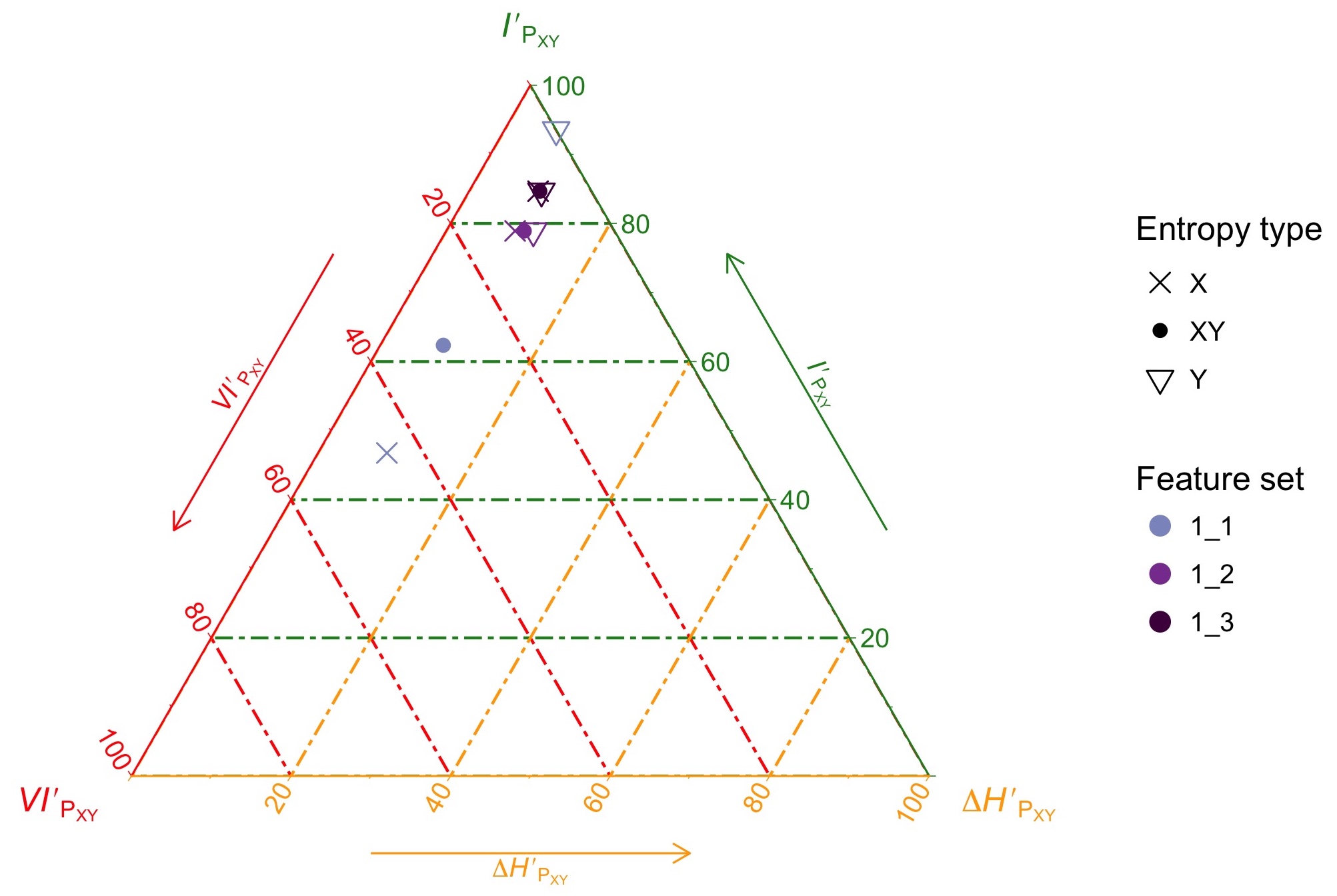}
		\label{fig:transfo:ICA:arthr}
	}
\caption{\textbf{(Color online) Split CMET exploration of feature transformation and selection with PCA (left) and ICA (right) on \emph{Iris}, \emph{Glass} and \emph{Arthritis} when selecting the first $n$ ranked features as obtained for each method.} The colors of the axes have been selected to match those of Figure \ref{fig:midiagrams}. 
\label{fig:transfo:independent}
}
\end{figure}

{ 
For instance, 
Figure~\ref{fig:transfo:independent}.\subref{fig:transfo:PCA:iris} presents the results of the PCA transformation on the logarithm of the features of Anderson's \texttt{Iris}. Crosses represent the information decomposition of the input features $\overline X$ using \eqref{eq:cmet:split:simplex:X} while circles represent the information decomposition of transformed features $\overline Y_i$ using \eqref{eq:cmet:split:simplex:Y} and filled circles the aggregate decomposition of \eqref{eq:cmet:simplex}. 
We represent several possible features sets $\overline Y_i$ as output where each is obtained selecting the first $i$ features in the ranking provided by PCA. For example, since \texttt{Iris} has four features we can make four different feature sets of $1$ to $i$ features, named in the Figure as ``1\_$i$'', that is, ``1\_1'' to ``1\_4''. 
The figure then explores how the information in the whole database $\overline X$ is transported to different, nested candidate feature sets $\overline Y_i$ as per the PCA recipe: choose as many ranked features as required to increase the transmitted information. 
} 
%

We first notice that all the points for $\overline X$ lie on a line parallel to the left side of the triangle and their average transmitted information is increasing, parallel to a decrease in remanent information. Indeed, the {redundancy} $\Delta H'_{\overline X}=\frac{\Delta H_{\overline X}}{H_{U_{\overline X}}}$ is the same regardless of the choice of $\overline Y_i$. The monotonic increase with the number of features selected $i$ in \emph{average transmitted information} $I'_{P_{\overline X \overline Y_i}} = \frac{I_{P_{\overline X \overline Y_i}}}{H_{U_{\overline X}}}$ in \eqref{eq:cmet:split:simplex:X} corresponds to the monotonic increase in absolute transmitted information $I_{P_{\overline X \overline Y_i}}$: 
for a given input set of features $\overline X$, the more output features are selected, the higher the mutual information between input and output. 
This is the basis of the effectiveness of the feature-selection procedure.

Regarding the points for $\overline Y_i$, note that the \emph{absolute} transmitted information also appears in the \emph{average}  transmitted information (with respect to $\overline Y_i$) as $I'_{P_{\overline X \overline Y_i}} = \frac{I_{P_{\overline X \overline Y_i}}}{H_{U_{\overline Y_i}}}$ in \eqref{eq:cmet:split:simplex:Y}. While  $I_{P_{\overline X \overline Y_i}}$ increases with $i$, as mentioned, we actually see a monotonic \emph{decrease}  in $I'_{P_{\overline X \overline Y_i}}$. 
The reason for this is the rapidly increasing value of the denominator $H_{U_{\overline Y_i}}$ as we select more and more features. 

{ 
Finally, notice how these two tendencies are conflated in the aggregate plot for the $\overline X \overline Y_i$ in Figure~\ref{fig:transfo:comparison}.\subref{fig:transfo:comparison:iris} that shows a lopsided,  inverted U pattern, peaking before $i$ reaches its maximum. This suggests that if we balance aggregated transmitted information against number of features selected---the complexity of the representation---in the search for a \emph{faithful} representation, the average transmitted information is the quantity to optimize, that is, the \emph{mutual determination} between the two feature sets. 
%

Figure~\ref{fig:transfo:independent}.\subref{fig:transfo:ICA:iris} presents similar results on the ICA transformation on the logarithm of the features of Anderson's \texttt{Iris} with the same glyph convention as before, but with a ranking resulting from carrying the ICA method \emph{in full} for each value of $i$. That is, we first work out $\overline Y_1$ which is a single component, then we calculate $\overline Y_2$ which the two best ICA components, and so on. The reason for this is that ICA does not rank the features it produces, so we have to create this ranking by carrying the ICA algorithm for all values of $i$ to obtain each $\overline Y_i$. 
Note that the transformed features produce by PCA and ICA are, in principle, very different, but the phenomena described for PCA are also apparent here: an increase in \emph{aggregate} transmitted information, checked by the increase of the denominator represented by $H_{U_{\overline Y_i}}$ which implies a decreasing \emph{average} transmitted information for $\overline Y_i$. 

With the present framework the question of which transformation is ``better'' for this dataset can be given content and rephrased as \emph{which transformation transmits more information on average on this dataset},  
and also, importantly, \emph{whether the aggregate information available in the dataset is being transmitted} by either of these methods. 
This is explored in Figure~\ref{fig:transfo:comparison} for \emph{Iris}, \emph{Glass} and \emph{Arthritis}, where, for reference, we have included a point for the (deterministic) transformation of the logarithm, the cross, giving an idea of what a lossless information transformation can achieve. 
\begin{figure}
	\centering	
	\subfloat[Comparing the transformations on \emph{Iris}]{
		\includegraphics[width=0.5\linewidth]{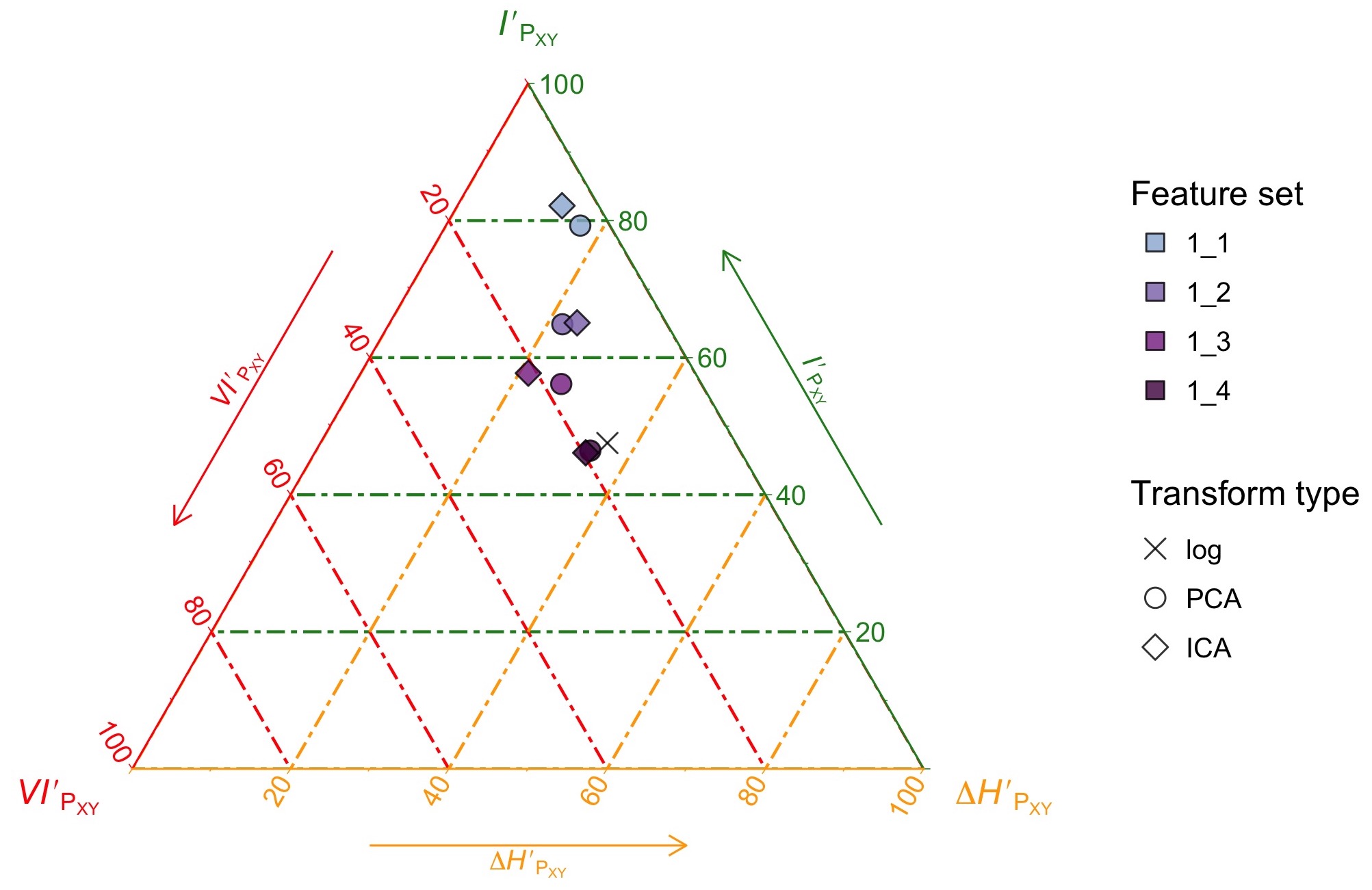}
			\label{fig:transfo:comparison:iris}
	}
	\subfloat[Comparing the transformations on \emph{Glass}]{
		\includegraphics[width=0.5\linewidth]{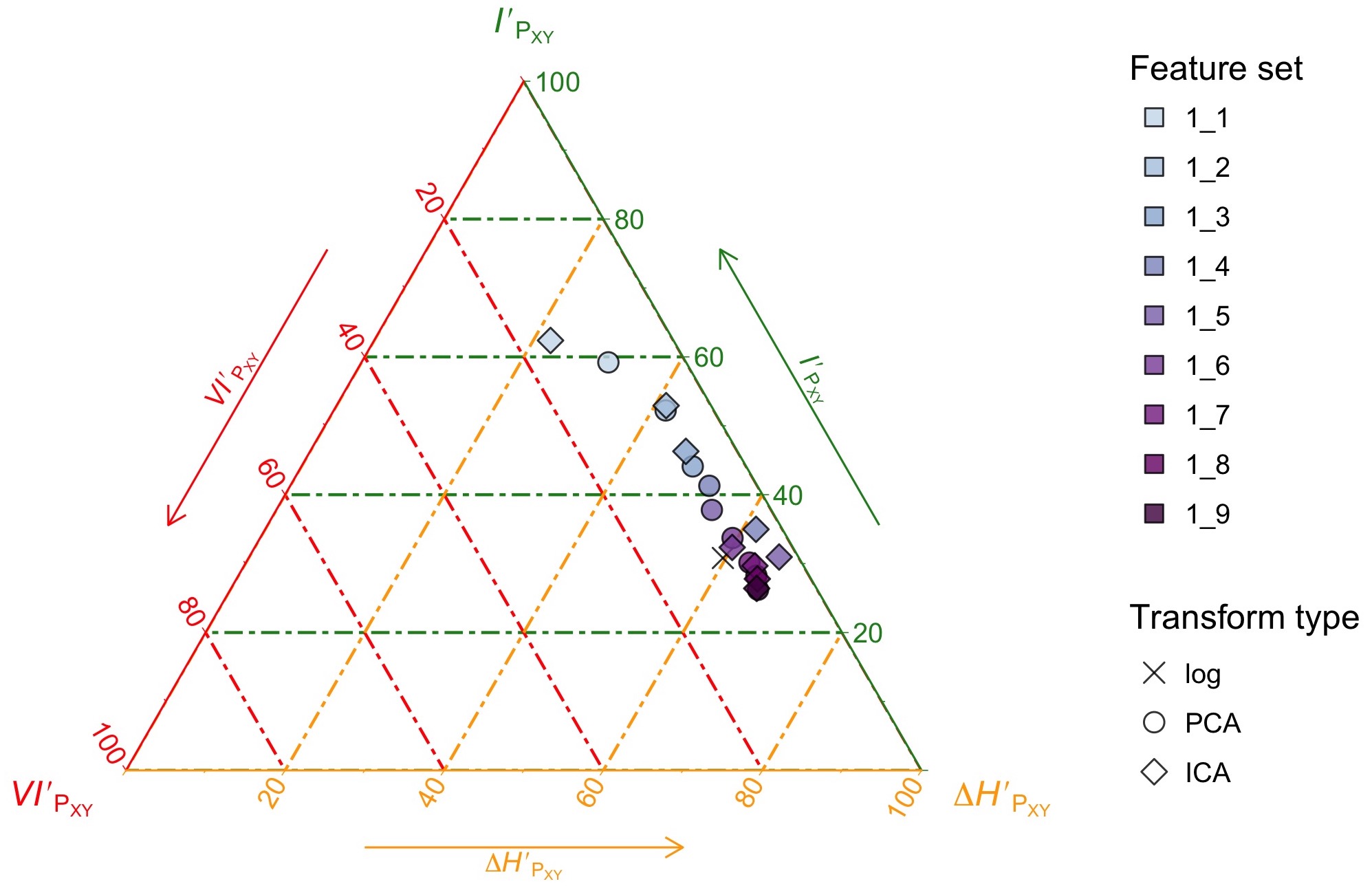}
			\label{fig:transfo:comparison:glass}
	}
	\\
	\subfloat[Comparing the transformations on \emph{Arthritis}]{
		\includegraphics[width=0.5\linewidth]{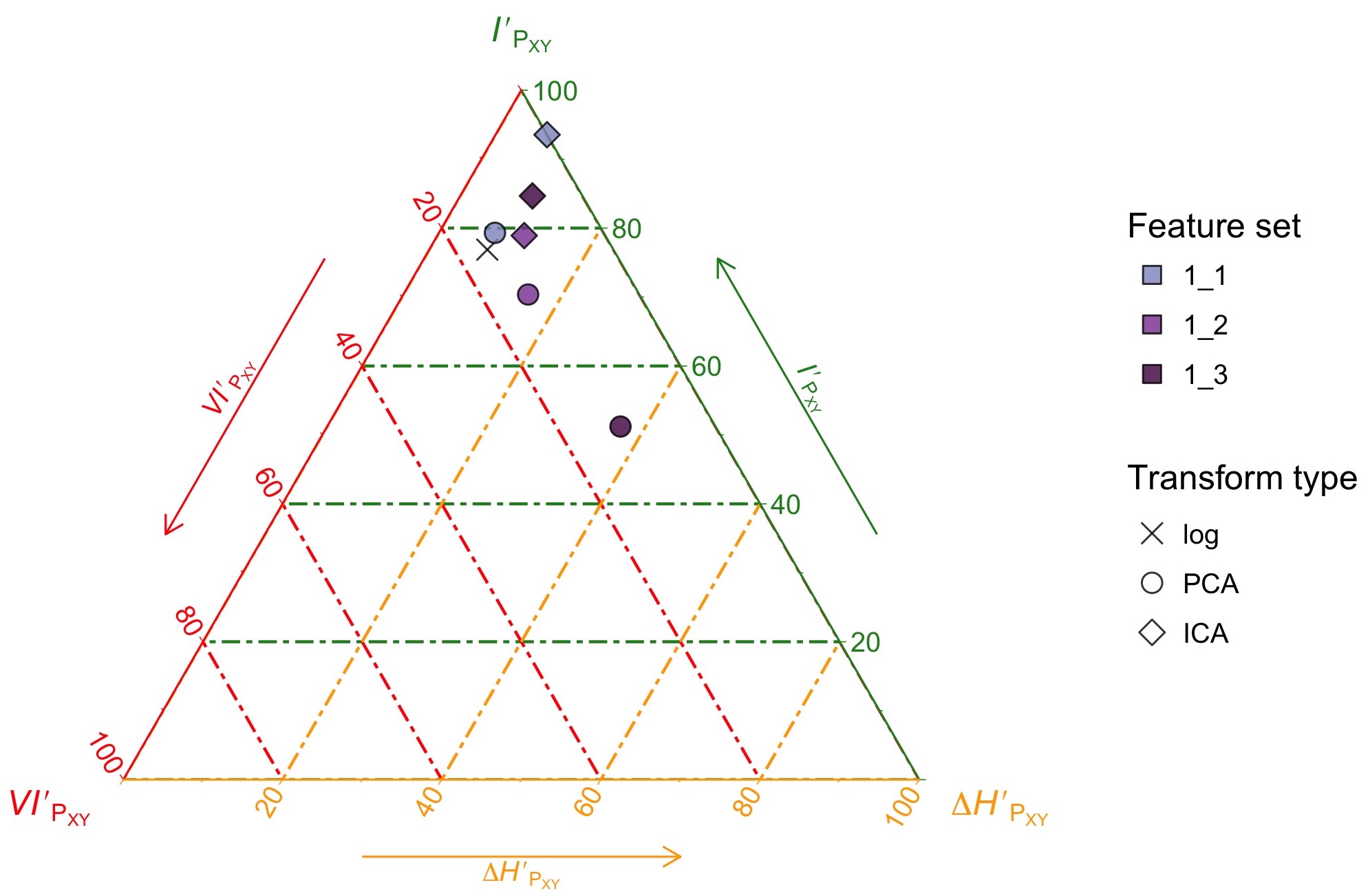}
			\label{fig:transfo:comparison:arthr}
	}
\caption{\textbf{(Color online) Comparison of PCA and ICA as data transformations using the CMET on \emph{Iris}, \emph{Glass} and \emph{Arthritis}. }
Note that these are the same positions represented as inverted triangles in Figures~\ref{fig:transfo:independent}.\protect\subref{fig:transfo:PCA:iris} and ~\ref{fig:transfo:independent}.\protect\subref{fig:transfo:ICA:iris}.
\label{fig:transfo:comparison}
}
\end{figure}

Consider Figure~\ref{fig:transfo:comparison}.\subref{fig:transfo:comparison:iris}  for \emph{Iris}.  The first interesting observation is that neither technique is transmitting all of the information in the database, which can be gleaned from the fact that both feature sets ``1\_4''---when all the features available have been selected---are below the cross. 
This clearly follows the data processing inequality, but is still surprising since transformations like ICA and PCA are extensively used and considered to work well in practice. 
In this instance it can only be explained by the advantages of the dimensionality reduction achieved.
Actually, the observation in the CMET suggests that we can \emph{improve on the average transmitted information per feature} by retaining the three first features for each PCA and ICA. 

The analysis of \emph{Iris} turns out to be an intermediate case between that of \emph{Arthritis} and \emph{Glass}, the latter being the most typical in our analysis.
This is the case with a lot of original features $\overline X$ which transmit very little private, distinctive information. 
The typical behavior, both for PCA and ICA is to select at first, features that carry very little average information $\overline Y_1$. As we select more and more transformed features, information accumulates but at a very slow pace as shown in Figures~\ref{fig:transfo:independent}.\subref{fig:transfo:PCA:glass} and~\ref{fig:transfo:independent}.\subref{fig:transfo:ICA:glass}. 
Typically, the transformed features chosen last are very redundant. In the case of \emph{Glass}, specifically, there is no point in retaining features beyond the sixth (out of 9) for either PCA or ICA as shown in Figure~\ref{fig:transfo:comparison}.\subref{fig:transfo:comparison:glass}.
As to comparing the techniques, in some similarly-behaving datasets PCA is better, while in others ICA is. In the case of \emph{Glass}, it is better to use ICA when retaining up to two transformed features, but it is better to use PCA when retaining between 2 and 6. 

The case of \emph{Arthritis} is quite different, perhaps due to the small number of original features $n=3$. 
Our analyses show that just choosing the first ICA component $\overline Y_1$---perhaps the first two---provides an excellent characterization of the dataset, being extremely efficient in what regards information transmission. 
This phenomenon is also seen in the first PCA component, but is lost as we aggregate more PCA components. Crucially, taking the $3$ ICA components amounts to taking all of the original information in the dataset, while taking the $3$ components in the case of PCA is rather inefficient, as confirmed by Figure~\ref{fig:transfo:comparison}.\subref{fig:transfo:comparison:arthr}.

All in all, our analyses show that the balance equations and entropy triangles are effective tools to visualize and assess the unsupervised transformation and selection of features in datasets. And that  this can be assessed from the information-theoretical heuristic of trying to maximize the average mutual information accumulated by the transformed features. 
} 


%% file: discussionCMET_R1.tex
The development of the multivariate case is quite parallel to the bivariate case. An important point to realize is that the multivariate transmitted information between two different random vectors $I_{P_{\overline X \overline Y}}$ is the proper generalization for the usual mutual information $I_{P_{XY}}$ in the bivariate case, rather than the more complex alternatives used in multivariate sources (see Section~\ref{sec:circa:MI} and ~\cite{tim:alf:fle:beg:14,val:pel:17b}). 
Indeed properties \eqref{eq:bi:internal} and \eqref{eq:bi:external} are crucial in transporting the structure and intuitions built from the bivariate channel entropy triangle to the multivariate one, of which the former is a proper instance. 
This was not the case with balance equations and entropy triangles for stochastic sources of information~\cite{val:pel:17b}. 

The crucial quantities in the balance equation and the triangle have been independently motivated in other works. 
First, multivariate mutual information is fundamental in Information Theory, and we have already mentioned the redundancy $\Delta H_{P_X}$~\cite{kay:03}. 
We also mentioned the input-entropy normalized $I'_{P_{\overline X\overline Y}}$ used as a standalone assessment measure in intrusion detection~\cite{gu:etal:06}.
Perhaps the least known quantity in the paper was the variation of information. Despite being inspired by the concept proposed by Meila~\cite{mei:07},   to the best of our knowledge it is completely new in the multivariate setting. However, the underlying concepts of conditional or remanent entropies have proven  their usefulness time and again. 
All of the above is indirect proof that the quantities studied in this paper are significant, and the existence of a balance equation binding them together important. 

The paragraph above notwithstanding, there are researchers who claim that Shannon-type relations cannot capture all the dependencies inside multivariate random vectors~\cite{jam:cru:17}. Due to the novelty of that work, it is not clear how much the ``standard'' theory of Shannon measures would have to change to accommodate the objections raised to it in that respect. But this question seems to be off the mark for our purposes: the framework of channel balance equations and entropy triangles has not been developed to look into the question of dependency, but of \emph{aggregate information transfer},  wherever that information comes from.  It may be relevant to source balance equations and triangles~\cite{val:pel:17b}---which have a different purpose---but that still has to be researched into.

The normalizations involved in \eqref{eq:cbet:simplex} and \eqref{eq:cmet:simplex}---respectively,  \eqref{eq:cbet:split:simplex}, \eqref{eq:cmet:split:simplex:X} and \eqref{eq:cmet:split:simplex:Y}---are similar conceptually: to divide by the logarithm of the total size of the domains involved whether it is the size of $X \times Y$ or that of $\overline X \times \overline Y$\,. 
Notice, first, that this is the same as taking the logarithm base these sizes in the non-normalized equations. The resulting units would not be bits for the multivariate case proper, since the size of $\overline X$ or $\overline Y$ is at least $ 2 \times 2 = 4$. 
But since the entropy triangles represent compositions~\cite{paw:ego:tol:15}, which are inherently dimensionless, this allows us to represent many different, and otherwise incomparable systems, e.g. univariate and multivariate ones with the same kind of diagram. 
Second, 
this type of normalization allows for an interpretation of the extension of these measures to the continuous case as a limit in the process of equipartitioning a compact support, as done, for instance, for the R\'enyi entropy in~\citep[\S~3]{jiz:ari:04} which is known to be a generalization of Shannon's. 
There are hopes, then for a continuous version of the balance equations for Renyi's entropy. 

Finally, note that the application presented in Section~\ref{sec:app} above, although principled in the framework presented here, is not conclusive on the quality of the analyzed transformations in general but only as applied to the particular dataset. For that, a wider selection of data transformation approaches, and many more datasets should be assessed. Furthermore, the feature selection process used the ``filter'' approach which for supervised tasks 
seems suboptimal. Future work will address this issue as well as how the technique developed here relates to the end-to-end assessment presented in~\cite{val:pel:14a} and the source characterization technique of~\cite{val:pel:17b}.

%
%

%% file: conclusionsCMET.tex
{ 
In this paper we have introduced a new way to assess quantitatively and visually the transfer of information from a multivariate source $\overline X$ to a multivariate sink of information $\overline Y$, using a heretofore unknown decomposition of the entropies around the joint distribution $P_{\overline X \overline Y}$.
%
For that purpose we have generalized a similar previous theory and visualization tools for bivariate sources greatly extending the applicability of the results:
\begin{itemize}
\item We have been able to decompose the information of a random multivariate source into three components a) the non-transferable divergence from uniformity $H_{P_{\overline X \overline Y}}$ which is an entropy ``missing'' in $P_{\overline X \overline Y}$, b) a  transferable but not transferred part, the variation of information $VI_{P_{\overline X \overline Y}}$, and c) the transferable and transferred information $I_{P_{\overline X \overline Y}}$ which is a known----but never considered in this context---generalization of bivariate mutual information.

\item Using the same principles as in previous developments, we have been able to obtain a new type of visualization diagram for this balance of information using de Finetti's ternary diagrams, which is actually an Exploratory Data Analysis tool. 
\end{itemize}

We have also shown how
to apply these new theoretical developments and the visualization tools to the analysis of information transfer in unsupervised feature transformation and selection, an ubiquitous step in data analysis, and, specifically, to apply it to the analysis of PCA and ICA. 
%
We believe this is a fruitful approach e.g. for the assessment of learning systems and foresee a bevy of applications to come. 
Further conclusions on this issue are left for  a more thorough later investigation. 
}

%% file: CMET-mdpi.bbl
\begin{thebibliography}{-------}
\providecommand{\natexlab}[1]{#1}

\bibitem[Goodfellow \em{et~al.}(2016)Goodfellow, Bengio, and
  Courville]{goo:ben:cou:16}
Goodfellow, I.; Bengio, Y.; Courville, A.
\newblock {\em {Deep Learning}}; MIT Press,  2016.

\bibitem[Shwartz-Ziv and Tishby(2017)]{sch:tis:17}
Shwartz-Ziv, R.; Tishby, N.
\newblock {Opening the Black Box of Deep Neural Networks via Information.}
\newblock {\em arXiv} {\bf 2017}, {\em 1703.00810 [cs.LG]}.

\bibitem[Tishby and Zaslavsky(2015)]{tis:zas:15}
Tishby, N.; Zaslavsky, N.
\newblock {Deep Learning and the Information Bottleneck Principle.}
\newblock  IEEE 2015 Information Theory Workshop,  2015.

\bibitem[Valverde-Albacete and Pel\'aez-Moreno(2014)]{val:pel:14a}
Valverde-Albacete, F.J.; Pel\'aez-Moreno, C.
\newblock 100\% classification accuracy considered harmful: the normalized
  information transfer factor explains the accuracy paradox.
\newblock {\em PLOS ONE} {\bf 2014}, pp. 1--10.
\newblock
  doi:{\changeurlcolor{black}\href{https://doi.org/10.1371/journal.pone.0084217}{\detokenize{10.1371/journal.pone.0084217}}}.

\bibitem[Valverde-Albacete and Pel\'aez-Moreno(2017)]{val:pel:17b}
Valverde-Albacete, F.J.; Pel\'aez-Moreno, C.
\newblock The Evaluation of Data Sources using Multivariate Entropy Tools.
\newblock {\em Expert Systems with Applications} {\bf 2017}, {\em
  78},~145--157.
\newblock
  doi:{\changeurlcolor{black}\href{https://doi.org/10.1016/j.eswa.2017.02.010}{\detokenize{10.1016/j.eswa.2017.02.010}}}.

\bibitem[Valverde-Albacete and Pel\'aez-Moreno(2010)]{val:pel:10b}
Valverde-Albacete, F.J.; Pel\'aez-Moreno, C.
\newblock Two information-theoretic tools to assess the performance of
  multi-class classifiers.
\newblock {\em Pattern Recognition Letters} {\bf 2010}, {\em 31},~1665--1671.

\bibitem[Yeung(1991)]{yeu:91}
Yeung, R.
\newblock {A new outlook on Shannon's information measures}.
\newblock {\em IEEE Transactions on Information Theory} {\bf 1991}, {\em
  37},~466--474.

\bibitem[Reza(1961)]{rez:61}
Reza, F.M.
\newblock {\em {An introduction to information theory}}; McGraw-Hill Electrical
  and Electronic Engineering Series, McGraw-Hill Book Co., Inc., New
  York-Toronto-London,  1961.

\bibitem[MacKay(2003)]{kay:03}
MacKay, D.J.C.
\newblock {\em {Information Theory, Inference and Learning Algorithms}};
  Cambridge University Press,  2003.

\bibitem[Shannon(1948{\natexlab{a}})]{sha:48a}
Shannon, C.E.
\newblock {A mathematical theory of Communication}.
\newblock {\em The Bell System Technical Journal} {\bf 1948}, {\em
  XXVII},~379--423.

\bibitem[Shannon(1948{\natexlab{b}})]{sha:48b}
Shannon, C.E.
\newblock {A mathematical theory of communication}.
\newblock {\em The Bell System Technical Journal} {\bf 1948}, {\em
  XXVII},~623--656.

\bibitem[Meila(2007)]{mei:07}
Meila, M.
\newblock Comparing clusterings---an information based distance.
\newblock {\em Journal of Multivariate Analysis} {\bf 2007}, {\em
  28},~875--893.

\bibitem[Pawlowsky-Glahn \em{et~al.}(2015)Pawlowsky-Glahn, Egozcue, and
  Tolosana-Delgado]{paw:ego:tol:15}
Pawlowsky-Glahn, V.; Egozcue, J.J.; Tolosana-Delgado, R.
\newblock {\em {Modeling and Analysis of Compositional Data}};
  Pawlowsky-Glahn/Modelling and Analysis of Compositional Data, John Wiley {\&}
  Sons: Chichester, UK,  2015.

\bibitem[Valverde-Albacete \em{et~al.}(2013)Valverde-Albacete, de~Albornoz, and
  Pel\'aez-Moreno]{val:car:pel:13}
Valverde-Albacete, F.J.; de~Albornoz, J.C.; Pel\'aez-Moreno, C.
\newblock A Proposal for New Evaluation Metrics and Result Visualization
  Technique for Sentiment Analysis Tasks.
\newblock  Information Access Evaluation. Multilinguality, Multimodality and
  Visualization. Proceedings of CLEF 2013; Forner, P.; henning M{\"u}ller.;
  Paredes, R.; Rosso, P.; Stein, B., Eds. Springer,  2013, Vol. 8138, {\em
  LNCS}, pp. 41--52.

\bibitem[Timme \em{et~al.}(2014)Timme, Alford, Flecker, and
  Beggs]{tim:alf:fle:beg:14}
Timme, N.; Alford, W.; Flecker, B.; Beggs, J.M.
\newblock {Synergy, redundancy, and multivariate information measures: an
  experimentalist{\textquoteright}s perspective}.
\newblock {\em Journal of Computational Neuroscience} {\bf 2014}, {\em
  36},~119--140.

\bibitem[James \em{et~al.}(2011)James, Ellison, and
  Crutchfield]{jam:ell:cru:11}
James, R.G.; Ellison, C.J.; Crutchfield, J.P.
\newblock {Anatomy of a bit: Information in a time series observation.}
\newblock {\em Chaos} {\bf 2011}, {\em 21},~037109--037109.

\bibitem[Watanabe(1960)]{wat:60}
Watanabe, S.
\newblock {Information theoretical analysis of multivariate correlation}.
\newblock {\em International Business Machines Corporation. Journal of Research
  and Development} {\bf 1960}, {\em 4},~66--82.

\bibitem[Tononi \em{et~al.}(1994)Tononi, Sporns, and Edelman]{ton:spo:ede:94}
Tononi, G.; Sporns, O.; Edelman, G.M.
\newblock {A measure for brain complexity: relating functional segregation and
  integration in the nervous system.}
\newblock {\em Proceedings of the National Academy of Sciences of the United
  States of America} {\bf 1994}, {\em 91},~5033--5037.

\bibitem[Studen{\'{y}} and Vejnarov{\'a}(1998)]{stu:vej:98}
Studen{\'{y}}, M.; Vejnarov{\'a}, J.
\newblock {The Multiinformation Function as a Tool for Measuring Stochastic
  Dependence}. In {\em Learning in Graphical Models}; Springer Netherlands:
  Dordrecht,  1998; pp. 261--297.

\bibitem[Han(1978)]{han:78}
Han, T.S.
\newblock {Nonnegative entropy measures of multivariate symmetric
  correlations}.
\newblock {\em Information and Control} {\bf 1978}, {\em 36},~133--156.

\bibitem[Abdallah and Plumbley(2012)]{abd:plu:12}
Abdallah, S.A.; Plumbley, M.D.
\newblock {A measure of statistical complexity based on predictive information
  with application to finite spin systems}.
\newblock {\em Physics Letters A} {\bf 2012}, {\em 376},~275--281.

\bibitem[Tononi(1998)]{ton:ede:spo:98}
Tononi, G.
\newblock {Complexity and coherency: integrating information in the brain}.
\newblock {\em Trends in Cognitive Sciences} {\bf 1998}, {\em 2},~474--484.

\bibitem[McGill(1954)]{mcg:54}
McGill, W.J.
\newblock {Multivariate information transmission}.
\newblock {\em Psychometrika} {\bf 1954}, {\em 19},~97--116.

\bibitem[Sun~Han(1980)]{han:80}
Sun~Han, T.
\newblock {Multiple mutual informations and multiple interactions in frequency
  data}.
\newblock {\em Information and Control} {\bf 1980}, {\em 46},~26--45.

\bibitem[Bell(2003)]{bell:03}
Bell, A.
\newblock {The co-information lattice}.
\newblock  Proceedings of the Fifth International Workshop on Independent
  Component Analysis and Blind Signal Separation; Murata, N.; Amari, S.i.;
  Cichocki, A.; Makino, S., Eds.,  2003.

\bibitem[Abdallah and Plumbley(2010)]{Abdallah:2010te}
Abdallah, S.A.; Plumbley, M.D.
\newblock {Predictive Information, Multiinformation and Binding Information}.
\newblock Technical Report C4DM-TR10-10, Queen Mary, University of London,
  2010.

\bibitem[Valverde~Albacete and Pel\'aez-Moreno(2016)]{val:pel:16a}
Valverde~Albacete, F.J.; Pel\'aez-Moreno, C.
\newblock The Multivariate Entropy Triangle and Applications.
\newblock  Hybrid Artificial Intelligence Systems (HAIS 2016), Proceedings;
  Springer: Seville (Spain),  2016; pp. 1--12.

\bibitem[Witten \em{et~al.}(2011)Witten, Eibe, and Hall]{wit:eib:hal:11}
Witten, I.H.; Eibe, F.; Hall, M.A.
\newblock {\em {Data mining. Practical machine learning tools and techniques}},
  3rd ed.; Morgan Kaufmann,  2011.

\bibitem[Pearson(1901)]{pea:01}
Pearson, K.
\newblock {On Lines and Planes of Closest Fit to Systems of Points in Space}.
\newblock {\em Philosophical Magazine} {\bf 1901}, pp. 559--572.

\bibitem[Bell and Sejnowski(1995)]{bel:sej:95}
Bell, A.J.; Sejnowski, T.J.
\newblock {An Information-Maximization Approach to Blind Separation and Blind
  Deconvolution}.
\newblock {\em Neural Computation} {\bf 1995}, {\em 7},~1129--1159.

\bibitem[Hyv{\"a}rinen and Oja(2000)]{hyv:oja:00}
Hyv{\"a}rinen, A.; Oja, E.
\newblock {Independent component analysis: algorithms and applications}.
\newblock {\em IEEE Transactions on Neural Networks} {\bf 2000}, {\em
  13},~411--430.

\bibitem[Bache and Lichman(2013)]{bac:lic:13}
Bache, K.; Lichman, M.
\newblock {UCI} Machine Learning Repository,  2013.

\bibitem[Gu \em{et~al.}(2006)Gu, Fogla, Dagon, Lee, and Skori\'{c}]{gu:etal:06}
Gu, G.; Fogla, P.; Dagon, D.; Lee, W.; Skori\'{c}, B.
\newblock Measuring Intrusion Detection Capability: An Information-theoretic
  Approach.
\newblock  Proceedings of the 2006 ACM Symposium on Information, Computer and
  Communications Security; ACM: New York, NY, USA,  2006; ASIACCS '06, pp.
  90--101.
\newblock
  doi:{\changeurlcolor{black}\href{https://doi.org/10.1145/1128817.1128834}{\detokenize{10.1145/1128817.1128834}}}.

\bibitem[James and Crutchfield(2017)]{jam:cru:17}
James, G.R.; Crutchfield, P.J.
\newblock {Multivariate Dependence beyond Shannon Information}.
\newblock {\em Entropy} {\bf 2017}, {\em 19},~531--545.

\bibitem[Jizba and Arimitsu(2004)]{jiz:ari:04}
Jizba, P.; Arimitsu, T.
\newblock {The world according to R{\'e}nyi: thermodynamics of multifractal
  systems}.
\newblock {\em Annals of Physics} {\bf 2004}, {\em 312},~17--59.

\end{thebibliography}
